# SAXS Investigation of Core-Shell Microgels with High Scattering Contrast Cores: Access to Structure Factor and Volume Fraction

Marco Hildebrandt,[1] Sergey Lazarev,[2] Javier Pérez,[3] Ivan A. Vartanyants,[2] Janne-Mieke Meijer, [4,*] Matthias Karg[1,*]

[1]Institut für Physikalische Chemie I: Kolloide und Nanooptik, Heinrich-Heine-Universität Düsseldorf, Universitätsstraße 1, D-40225 Düsseldorf, Germany

E-Mail: karg@hhu.de

[2]Deutsches Elektronen-Synchrotron DESY, Notkestraße 85, 22607 Hamburg, Germany

[3]Synchrotron SOLEIL, L'Orme des Merisiers, Saint-Aubin, BP 48, 91192 Gif-sur-Yvette Cedex, France

[4]Department of Applied Physics and Institute for Complex Molecular Systems, Eindhoven University of Technology, P.O. Box 513, 5600 MB Eindhoven, The Netherlands

E-Mail: j.m.meijer@tue.nl

**ABSTRACT:** To explore dense packings of soft colloids, scattering experiments are ideal to access the structure factor. However, for soft microgels determination of the structure factor is difficult because of the low contrast of the polymer network and potential microgel interpenetration and deformation that change the form factor contribution. Here, we employ small-angle X-ray scattering (SAXS) to study soft, thermoresponsive microgels with poly-*N*-isopropylacrylamide (PNIPAM) shells and gold nanoparticle cores. The scattering of the gold cores dominates the scattering patterns and allows precise determination of the microgel volume fraction over a broad range of concentrations. At high volume fractions we find distinct patterns with sharp Bragg peaks allowing extraction of the structure factor and characterization of the phases combined with UV-Vis spectroscopy. The unique scattering contrast of our core-shell microgels combined with SAXS opens up new ways to investigate dense packings of soft microgels including in situ studies of phase transitions.

## Introduction

Microgels are composed of cross-linked polymer networks that are swollen by a solvent and possess a dual colloid-polymer nature.[1-3] Through the polymer composition, microgels can be rendered responsive towards various external stimuli such as pH, ionic strength and temperature.[4-10] The most prominent example for microgels that respond to changes in temperature undergoing a pronounced volume phase transition (VPT) is composed of poly-$N$-isopropylacrylamide (PNIPAM). PNIPAM microgels were first synthesized by Pelton and Chibante[11] and received great attention ever since for example as model system for soft colloids.[5, 12-15] Furthermore, such responsive microgels are of interest for different applications, for instance photonic crystals,[13] 3D bioprinting,[16] color-changing systems,[17, 18] as well as viscosity modifiers and lubricants.[19, 20]

Similar to hard spheres, microgels can form crystalline phases in 2D and 3D assemblies. The fact that the microgel size can be tuned by external parameters and thereby allowing for in-situ changes of the volume fraction, $\phi$, makes responsive microgels highly attractive for the study of crystallization and melting phenomena.[21, 22] Furthermore, the softness and deformability of microgels extend their phase diagram even above the hard sphere limit ($\phi = 0.74$). In 3D assemblies of hard spheres, the maximum packing density is reached when particles are in direct contact in close packed crystalline structures, namely fcc (face centred cubic) or hcp (hexagonally close packed). Higher packing fractions are not possible because hard spheres cannot be deformed and/or compressed. In contrast, upon contact, microgels can deform, shrink and/or interpenetrate resulting in apparent volume fractions above the hard sphere limit and even above unity, $\phi > 1$.[20] For the real volume fraction the changes in individual microgel volume at high packings needs to be consid-ered. The difference between the apparent and the real volume fraction of soft and deformable microgels has been addressed recently by Scotti et al.[23] Clearly, understanding the influence of the softness and responsiveness of microgels on the structure and dynamics in dense dispersions is important for their applications as well as of fundamental interest to study, for example, crystallization, melting as well as glass or jamming transitions.

For relatively large microgels with sizes in the micron range, optical microscopy techniques have been used for investigating the local organization of microgel systems on a single particle level.[17, 24, 25] The overlapping of microgels in dense packings however limits the resolution in optical microscopy. In addition, different scattering methods, such as dynamics and static light scattering (DLS and SLS) or small-angle X-ray and neutron scattering (SAXS and SANS) were frequently employed to study the bulk structure and dynamics of microgel systems – mostly focusing on microgels too small to be observed with optical microscopy.[26-28] A major advantage of scattering methods over microscopy methods is that much higher particle numbers can be addressed in one measurement giving access to ensemble averages with great statistics. At high volume fractions, in particular, where microgels form colloidal crystals, scattering methods are much more powerful to resolve the long-range 3D order. However, it remains difficult to unambiguously decouple the form factor, $P(q)$, and the structure factor, $S(q)$, from scattering profiles. On the one hand both contributions appear on similar length scales (or ranges of scattering vector, $q$), thus requiring the form factor to be determined with great statistics. On the other hand, in contrast to hard spheres, the microgel form factor can change with concentration, in particular for dense packings due to interpenetration and/or deformation.[29, 30] Furthermore, the form factor of microgels is much more complex than that of hard spheres due to the inhomogeneous distribution of the cross-linker leading to the well-known fuzzy sphere morphology[27, 31] and the overlap of their outer polymer segments at high concentrations.[32]

Recently, there has been an increased interest in microgels that contain metallic nanoparticles. Such hybrid microgels are attractive for the development of new materials with applications in photonics, plasmonic lasing, and opto-electronic devices.[22, 33-35] The interest has arisen because the nanoparticle core, for example silver or gold, can host localized surface plasmon resonances (LSPRs),[36, 37] while the interparticle spacing is controlled by the microgel shell that also governs the formation of 2D or 3D colloidal crystals.[22] For core-shell (CS) microgels consisting of gold nanoparticles (AuNPs) coated with a cross-linked PNIPAM shell it was shown that these behave as soft microgel systems, forming 2D and 3D crystal lattices.[38-42] The 2D lattices were found to show surface lattice resonances (SLR) when the inter-particle spacing was in the order of the visible wavelength leading to plasmonic-diffractive coupling.[35, 43-45] Due to the strong LSPR absorption of the AuNP cores, however, the use of optical microscopy methods is limited and so far the 3D crystal lattices have been mostly investigated with UV-Vis spectroscopy focusing on the Bragg reflection.[22] A recent study also explored SANS for the characterization of 3D colloidal crystals and found that the PNIPAM shell leads to the formation of a dominant fcc crystal lattice, irrespective of the presence of the small gold core.[40] However, in SANS the scattering arises from the polymer shell and therefore the same interferences between form and structure factor hampers the detailed analysis of the structure factor. It further remains unclear if the gold cores are truly located in the centers of microgel shells. Lapkin *et al*. recently studied a similar system focusing on the melting and crystallization of CS microgels at high volume fraction.[21] Thanks to the high scattering contrast cores, ultra small-angle X-ray scattering (USAXS) could be used to provide detailed information on the phase transition upon changes in temperature in situ. The appearance of sharp Bragg peaks in the crystalline regime allowed a detailed analysis of the melting and crystallization behavior.

In this work, we focus on studying PNIPAM microgels with small AuNP cores by SAXS. We exploit the large size and electron density difference between the cores and the swollen PNIPAM shells that will lead to the main scattering contribution arising from the AuNP cores. We investigate different low concentrations of the CS microgel system and demonstrate that the AuNP core allows for the precise determination of the apparent microgel volume fraction. We also explore the formation of colloidal crystals driven by the microgel shells and investigate a broad concentration series. We determine the phase behavior and interparticle spacing as a function of microgel volume fraction. Finally, we compare our findings from SAXS with results from Bragg peak analysis using UV-Vis spectroscopy.

## Experimental Methods

**Chemicals.** Gold(III) chloride trihydrate (HAuCl$_4$; Sigma-Aldrich, ≥99.9%), sodium citrate dihydrate (Sigma Aldrich, ≥99.0%), butenyl amine hydrochloride (Sigma Aldrich,97.0%), sodium dodecyl sulfate (SDS; Merck, ≥95.0%), $N$-isopropylacrylamide (NIPAM; TCI 98.0%), $N,N'$-methylenebis(acrylamide) (BIS; Aldrich, 99%), potassium peroxodisulfate (PPS; Sigma ≥99.0%), sodium chloride (NaCl; Fischer Chemical,; Ph. Eur.). Water was purified with a Milli-Q System (Millipore), resulting in a final resistivity of 18 MΩ cm.

**Synthesis of Core-Shell Microgels.** CS microgels with spherical gold cores and polymer shells consisting of chemically cross-linked PNIPAM were synthesized via seeded precipitation polymerization.[46] NIPAM (1.169 g; 10.3 mmol) and BIS (0.239 g; 1.6 mmol) were dissolved in 600 mL of Milli-Q water followed by degassing with argon at a temperature of 70 °C for 1.5 h. Before initiating the reaction with 12 mg PPS dissolved in 1 mL of water, 8.4 mL of a stock dispersion of spherical Au-NP seeds[38] with a mean radius of 7.4 nm (from transmission electron microscopy) and an elemental gold concentration of $[Au^0]$ = 0.02 mol/L were added to the reaction mixture. After the polymerization was initiated, the reaction was continued for 4 h. The final CS microgels were purified by three centrifugation steps, each for 3 h at 7500 rcf and redispersion in water to remove residues of salt, unreacted monomer and potentially non-cross-linked polymer. Finally, the CS microgels were freeze-dried. We take into account for a residual water content of 5.7% in the final freeze-dried sample.[40, 47]

**Sample Preparation.** CS microgel dispersions with different weight concentrations were prepared by diluting a highly concentrated stock dispersion made from freeze-dried CS microgels. In this way sample dispersions within the concentration range from 0.5 wt% to 22.5 wt% were prepared. Although care was taken to control all wt% concentrations, for the very dense samples we sometimes observed CS particle deposition on the capillaries. This loss of particles in combination with solvent evaporation could lead to changes in the wt%.

**Transmission Electron Microscopy.** Transmission electron microscopy (TEM) was conducted with a JEOL JEM-2100Plus TEM in bright-field mode operated with an acceleration voltage of 80 kV. Samples were prepared via drop casting of a dilute aqueous CS microgel dispersion on carbon coated copper grids (200 mesh, Electron Microscopy Science). The grids were dried at room temperature for several hours before investigation. The particle size was determined from the TEM images using the GMS 3 software from Gatan as well as ImageJ.[48, 49]

**Dynamic Light Scattering.** Temperature dependent dynamic light scattering (DLS) was performed with a Malvern Zetasizer Nano S ($\lambda$ = 633 nm; $\theta$ = 173°). Three measurements at each temperature in a range of 15 to 65°C with a step of 1 K were taken with acquisition times of 60 s each. Dilute aqueous microgel dispersions were measured in standard polystyrene cuvettes with 1 cm pathlength. Hydrodynamic radii $R_h$ (z-average) were determined with cumulant analysis provided by the instrument software.

**Electrophoretic Mobility Determination.** Electrophoretic mobility was measured with a Malvern Zetasizer Nano Z ($\lambda$ = 633 nm; $\theta$ = 173°) at a temperature of 20°C. The CS microgels were dispersed in $10^{-4}$ M aqueous NaCl solution to provide a constant ionic background.

**UV-Vis Absorbance Spectroscopy**. UV-Vis spectra were recorded using a SPECORD S 600 (Analytik Jena) UV-Vis spectrophotometer. Dilute samples were measured in 1 cm PMMA cuvettes. Dense samples at high volume fractions were measured in 0.2 mm × 4.0 mm × 50 mm capillaries (VitroTubes). Measurements were performed at room temperature.

**Small-Angle X-Ray Scattering.** In-house SAXS measurements were performed on a Xeuss 2.0 (XENOCS) equipped with an X-ray beam of 8.048 keV, a sample to detector distance of 1 m and an acquisition time of 3600 s. Scattering patterns were collected with a Pilatus3R 300K with an area of 83.8 x 106.5 mm$^2$ and a pixel size of 172 x 172 µm$^2$. This setup provides a $q$-range of 0.03 nm$^{-1}$ < $q$ < 3.5 nm$^{-1}$. We want to mention that the resolution in the $q$-range of 0.03 – 0.1 nm$^{-1}$ is much lower compared to synchrotron SAXS. Dilute samples were measured in 1 mm round capillaries (WJM Glas) at a temperature of 20 °C. The measured signal was background corrected for the scattering of water, normalized to absolute scale using the thickness of the capillary and the scattering of glassy

carbon as reference and finally radially averaged with the Foxtrot software provided by Xenocs.[50]

Synchrotron SAXS measurements on dense samples at high volume fractions were performed at the SWING beamline at the SOLEIL synchrotron in Saint-Aubin (France). An X-ray beam of 8 keV ($\lambda$ = 0.155 nm) was employed with a sample to detector distance of 6497 mm. An Eiger 4M detector with an area of 155.2 × 165.5 mm$^2$ and a pixel size of 75 × 75 µm$^2$ was used to collect the 2D scattering patterns. This setup provides a $q$-range of 0.01 nm$^{-1}$ < $q$ < 1.0 nm$^{-1}$. Samples with different wt% were prepared in 0.2 mm × 4.0 mm × 50 mm capillaries (VitroTubes). These thin-walled rectangular capillaries provide short optical paths and also allow for measurements by SAXS and by UV-Vis absorbance spectroscopy on the same sample. Due to the high viscosity of high wt% samples, samples were heated to approximately 50 °C, which is well above the volume phase transition temperature (VPTT) of the microgel shells. At this temperature the volume fraction is significantly reduced due to the shrinkage of the PNIPAM shells lowering also the dispersion viscosity. By applying a small reduced pressure to one opening of the capillaries, the dispersions were sucked inside the capillaries. The capillaries were sealed with two-component epoxy glue. Prior to investigation, all samples were annealed at a temperature of approximately 50 °C. Subsequently, the samples were slowly cooled to room temperature during at least 1 h. All samples were measured at a temperature of 20 °C with an acquisition time of 100 ms. Background corrections were performed on the recorded 2D SAXS patterns before analysis with the Foxtrot software provided by SOLEIL.[50]

Radially averaged scattering profiles were analysed with the SASfit software by Kohlbrecher.[51] The 2D SAXS patterns were analysed with the Software Scatter by Förster and Apostol.[52]

## Results and Discussion

**Characterization in the Dilute State**. Seeded precipitation polymerization was used to synthesize CS microgels that have high electron density AuNP cores and low electron density hydrogel shells composed of chemically cross-linked PNIPAM. A schematic depiction of the CS morphology is shown in **Figure 1a.** In order to verify this CS structure experimentally and to determine the yield of encapsulation, the CS microgels were studied by TEM. **Figure 1b** shows a representative TEM image where the CS structure with the high contrast, spherical AuNP cores and the low contrast PNIPAM shells can be clearly identified. Due to drying effects on the TEM grids and the high vacuum conditions during the measurements, the microgel shells are in a collapsed state with dimensions much smaller than in dispersion under good solvent conditions (red, dashed circle in the TEM image). We also attribute the drying effects to be responsible for the AuNP cores not appearing in the centre of the microgels in TEM images (see **Figure 1b**). In addition, some microgels without cores could be observed (see **Figure S1** in the Supporting Information). The percentage of microgels that do not feature a AuNP core is low however (< 1 %) in agreement to findings in a previous work.[46] The mean radius of the AuNP cores, $R_c$, was determined as $R_c$ = 6.6 ± 0.7 nm. A corresponding histogram of the core sizes determined by TEM analysis is shown in **Figure S2** in the Supporting Information. A UV-Vis extinction spectrum measured from a dilute dispersion of the AuNP cores (no shells) reveals a single dipolar LSPR with a resonance maximum at $\lambda_{LSPR}$ = 524 nm (see **Figure S3** in the Supporting Information). Due to the rather small size of the AuNPs the LSPR is related to absorption of light while scattering, that scales with the sixth power of the particle radius, is negligible. The extinction properties change significantly when the much larger PNIPAM shell is added. **Figure 1c** shows the corresponding extinction spectrum of the CS microgels measured from dilute dispersion. While the LSPR of the AuNP cores is still visible as a peak at 524 nm wavelength, an increased absorbance is observed at lower wavelengths due to Rayleigh-Debye-Gans scattering of the larger PNIPAM shells. This superposition of scattering from the PNIPAM shells and absorption from the AuNPs cores hampers the determination of particle number concentrations from extinction values.[47] We will later on show that SAXS is the ideal tool for extracting number concentrations with great precision. The total size of the CS microgels in dilute dispersion was determined by DLS. At 20 °C, i.e. in the swollen state, we measured a hydrodynamic radius, $R_h$, of 105 ± 1 nm. The VPT of the PNIPAM shell was studied using temperature-dependent DLS measurements. **Figure S4** in the Supporting Information shows the evolution of $R_h$ as a function of temperature. With increasing temperature $R_h$ decreases continuously until a plateau is reached when the shell is in a fully collapsed state at T > 50°C with $R_h$ approaching approximately 65 nm. The VPTT is found at approximately 36 °C which agrees well to other PNIPAM microgels with similar cross-linking densities.[40, 47] The finding that we observe a pronounced temperature response for such high nominal cross-linker contents is in agreement with a recent study by Ponomareva et al. where similar CS microgels with varying size and cross-linker contents were studied in detail by different methods including light and neutron scattering.[31] The electrophoretic mobility of the CS microgels was investigated because it is known that electrostatic interactions can have an influence on the interparticle interactions.[53, 54] We find an electrophoretic mobility of -1.6 μm cm/Vs at 20 °C for the CS microgels. This slightly negative charge is related to the anionic initiator used in the synthesis. Therefore, we consider our microgels to be weakly electrostatically stabilized.

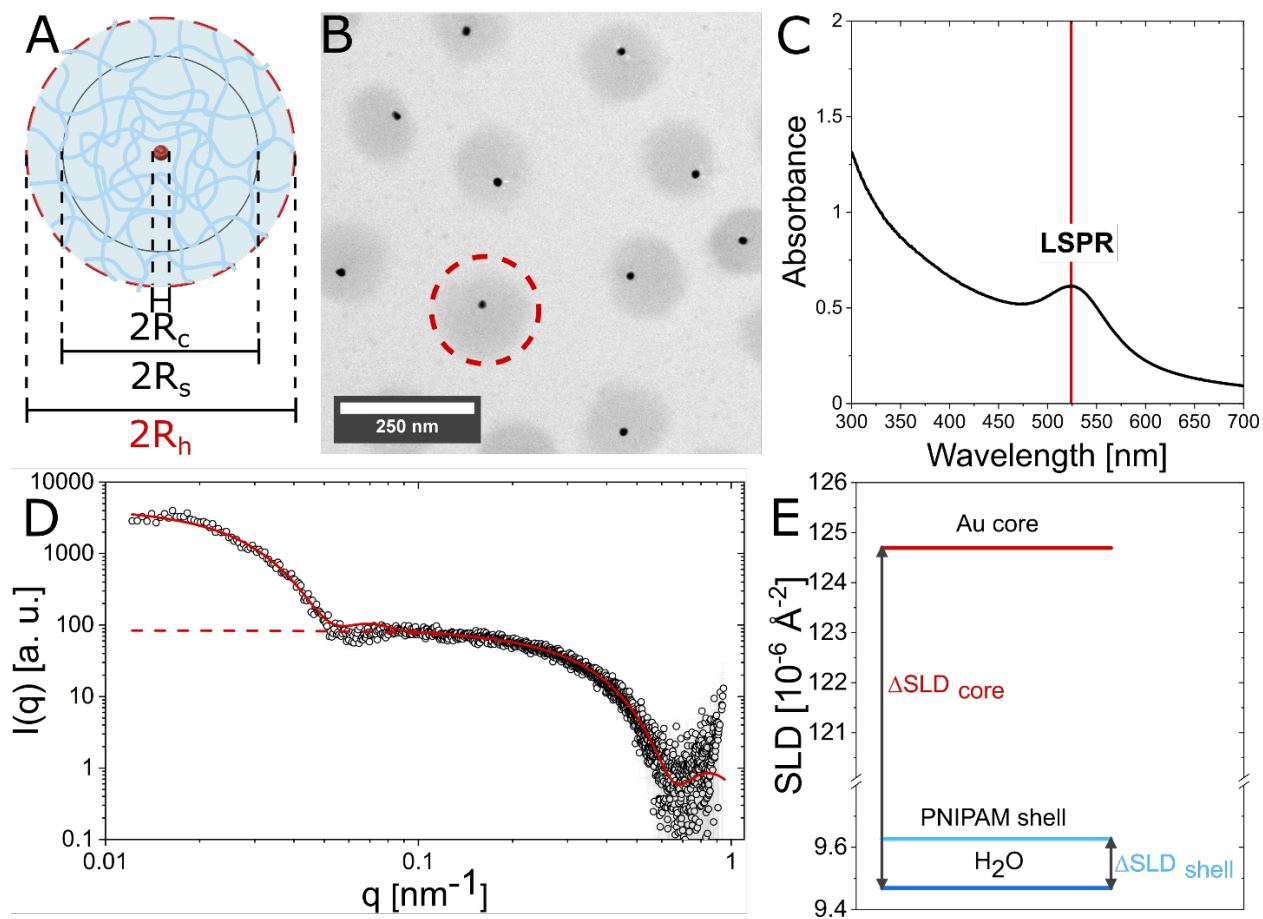

**Figure 1.** Characterization of Au-PNIPAM CS microgels. A) Schematic illustration of the CS structure and the most relevant radii accessible by scattering experiments with $R_c$ the radius of the AuNP core, $R_s$ the radius of the CS microgel from SAXS and $R_h$ the hydrodynamic radius. B) Representative TEM image of the CS microgels. The red circle indicates the dimensions of the swollen PNIPAM shell as obtained from DLS. C) UV-Vis absorbance spectrum of a dilute dispersion of CS microgels. The red line highlights the position of the LSPR. D) Synchrotron SAXS profile from a dilute dispersion (0.5 wt%) of the CS microgels recorded at 20 °C. The dashed line corresponds to the form factor fit of the AuNP scattering contribution. The solid line corresponds to a form factor model that combines two polydisperse hard spheres with different size and different scattering contrast. E) Scattering length density differences between Au, swollen PNIPAM microgel shell (80 % water content) and water as the dispersion medium. Note the break in scale due to the large difference between core and shell SLD. SLDs were obtained from the SLD calculator provided by NIST.[55]

To determine the form factor of the CS microgels and in particular to determine the contributions of core and shell scattering, we studied dilute microgel dispersions by SAXS. **Figure 1d** shows the radially averaged synchrotron SAXS profile measured at 20 °C. The scattered intensity $I(q)$ depends on the scattering contrast ($\Delta SLD$) given by the scattering length density (SLD) for homogenous scattering objects, the particle number density $N$,

the scattering object volume $V_P$, the form factor $P(q)$ and the structure factor $S(q)$, according to

$$I(q) = N(\Delta SLD)^2 V_p^2 P(q) S(q) \tag{1}$$

For dilute particle dispersions interparticle interactions can be considered negligible and $S(q) \approx 1$. In the mid to high $q$-region ($q > 0.1$ nm$^{-1}$) the SAXS profile is dominated by the form factor contribution of the AuNP cores. This contribution can be described by a simple polydisperse sphere model as the dashed red line in **Figure 1d** illustrates. In the following we will focus on the analysis of the core form factor and its forward scattering intensity. However, in the scattering profile we also observe an increase in scattering intensity in the low $q$-region ($q < 0.1$ nm$^{-1}$) that is related to scattering from the swollen PNIPAM shell. The contribution of the core is more distinct due to the difference in the SLD of the AuNP core and the PNIPAM shell as shown in **Figure 1e.** The difference in SLD of the AuNP cores and water exceeds that of the swollen PNIPAM shells with respect to water by almost three orders of magnitude, explaining the dominant contribution of the core in the mid to high $q$-region. The water content of PNIPAM microgels with different cross-linker densities up to 25 mol% (nominal) was investigated in detail by using SANS recently.[31] Even for such high nominal cross-linker amounts water contents of more than 80% in the swollen and more than 60% in the collapsed state were found. However, even though the shell possesses a lower contrast, it still contributes to the scattering profile at low $q$ due to its much larger volume. As a simple approach to describe the core and the shell contribution we use the simple sum of two polydisperse spheres taking into account the difference in scattering contrast of the cores and shells (solid red line **Figure 1d**, for details see Supporting Information). Qualitatively this simple model describes the measured data suf-ficiently well. We want to note that a more complex form factor model that considers interference between core and shell and also accounts for the inhomogeneous cross-linker distribution in the shell would be more appropriate.[21, 56] However, due to the rather small scattering contribution from the shell and the absence of pronounced form factor oscillations in the accessible $q$-range, we do not want to go deeper into such an analysis here. In the Supporting Information a form factor analysis comparing different models used to fit the scattering profile of the dilute CS microgel sample is given in **Figure S5**. In the following, the scattering signal from the cores only is what we will be using for further analyses in this study. From the fit of the core scattering contribution we find $R_c$ = 6.5 ± 0.6 nm which is slightly smaller than the radius obtained from TEM (6.6 ± 0.7 nm). This difference can be explained by approximation of a perfect sphere in SAXS while we measure slightly anisotropic shapes in TEM. The form factor fit resulting from the linear combination of two polydisperse spheres describes the full scattering profile sufficiently well and we determine a radius of $R_s$ = 77.9 ± 8.7 nm for the CS microgel. The reason $R_s$ is significantly smaller than $R_h$ is that the model does not take into account the fuzziness of the PNIPAM shell, while the outer low cross-linked region still contributes to the hydrodynamic dimensions. It is well known that PNIPAM microgels possess a gradient in cross-linker density due to the faster consumption of the cross-linker BIS during the precipitation polymerization.[57, 58] This gradient also leads to dangling end chains at the surface that contribute to $R_h$ but are not resolved in SAXS due to their very low contrast.[6, 29, 59] This is also in good agreement with our previous results.[21, 40] Also, the radius of gyration $R_g$ = 71.1 ± 1.7 nm determined from Guinier analysis is smaller than $R_s$ (see **Figure S6** and discussion in the Supporting Information).This can also be ascribed to the gradient in cross-linker density. A similar difference is found when we compare $R_g$ and $R_h$ where we find the ratio $R_g/R_h$ = 0.68, which is typically observed for microgels, for example studied by light scattering.[15]

**Volume Fraction.** Due to their soft and deformable nature as well as the dangling ends that allow for interpenetration in dense packings, the determination of the volume fraction, $\phi$, of microgel dispersions is challenging and often defective. One approach to extract volume fractions from dilute microgel dispersions is relative viscosimetry.[23, 60] Here, owed to the strong scattering contrast of the AuNP cores, we can use the absolute SAXS intensity to determine the number concentration, $N$ (see equation 1), of the CS microgels very precisely in the low concentration regime. **Figure 2a** shows radially averaged scattering profiles recorded for dilute CS microgel dispersions ($S(q) \approx 1$) at 20°C with known concentrations (wt%). The red solid lines correspond to the respective form factor fits of the AuNP core contribution. It is important to note that the scattering curves are not offset and the increase in $I(q)$ is a direct result of the increase in concentration and hence $N$. Because we normalized $I(q)$ to absolute intensity (cm$^{-1}$), the particle number density and hence $\phi_{core}$ can be determined from the intensity $I_0$ extrapolated to $q \rightarrow 0$ nm$^{-1}$ according to:

$$N = \frac{I_0 \, N_A \, \rho_{core}^2}{M_{core} \, m_{core} \, \Delta SLD^2} \tag{2}$$

With $\rho_{core}$ the AuNP core density, $N_A$ Avogadro's number, $M_{core}$ the molecular weight and $m_{core}$ the mass of the AuNP core, and $\Delta SLD$ the scattering length density differences between gold and water (for more details see Supporting Information). Once $N$ has been determined, the volume fraction of the CS particles, $\phi_{CS}$, can be obtained from the number concentration $N$ and the hydrodynamic radius $R_h$:

$$\phi_{CS} = N \, \frac{4}{3} \pi \, R_h^3 \tag{3}$$

We employ $R_h$ because Stieger *et al.*[27] showed that this is the most accurate size to describe soft microgels in the swollen state. The forward scattering intensities, $I_0$, were determined from linear extrapolation of the scattering data in a Guinier plot where only the $q$-range relevant to scattering from the AuNP cores is considered. The inset in **Figure 2a** shows the Guinier plot for the differently concentrated samples and the corresponding linear fits (red lines). More details about the fitting procedure and a larger scale image of the inset in **Figure 2a** are provided in the Supporting Information (**Table S3** and **Figure S7**). For the fits we have ignored the scattering from the shell, visible at $q < 0.06$ nm$^{-1}$.

**Figure 2b** shows the resulting relation between the sample concentration in wt% and the extracted volume fraction of CS microgels, $\phi_{cs}$, derived from $N$ resulting from the SAXS data and using the hydrodynamic radius ($R_h = 105$ nm) in the swollen state at 20 °C. As expected, there is a linear relation between $\phi_{cs}$ and the CS microgel concentration in wt%. The slope of the linear fit to the data provides the scaling between the two quantities and allows us the calculation of the volume fraction for any given concentration in wt% (see also **Figure S8** in the Supporting Information). We find:

$$\phi_{CS} = (0.130 \pm 0.001) \frac{\text{mass concentration in wt\%}}{\text{wt\%}} .  \tag{4}$$

A more detailed overview of concentrations and the respective volume fractions is presented in **Table S4** in the Supporting Information.

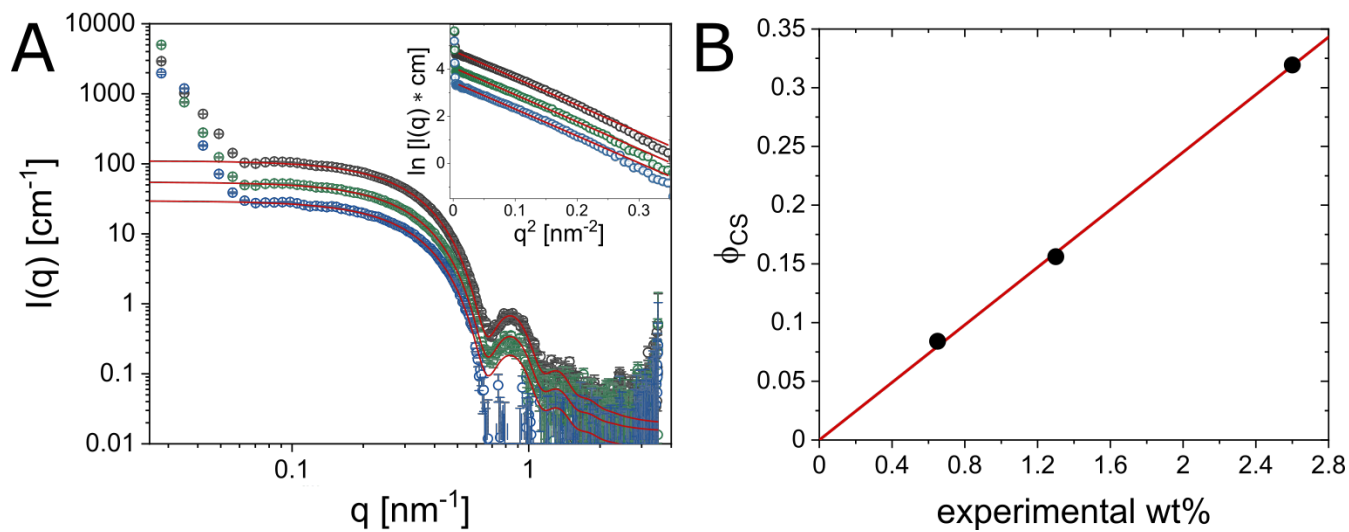

**Figure 2.** A) Radially averaged SAXS profiles for differently concentrated dispersions of the CS microgels in the dilute regime. The concentrations were 0.62 wt% (blue), 1.23 wt% (green) and 2.45 wt% (dark grey). Red lines correspond to the form factor fits only taking into account the scattering contribution of the AuNP cores. The inset shows the SAXS data in the Guinier plot representation with the red lines corresponding to linear fits used for the extrapolation to obtain $I_0$ at $q \rightarrow 0$ nm$^{-1}$. B) Determined volume fraction $\phi_{CS}$ for each experimental wt% (black circles) showing a linear relation (red line).

**Separation of the form factor and structure factor.** For dense packings of purely organic microgels without high contrast cores, form factor and structure factor typically overlap to some extend in the low $q$-range.[26, 27] Therefore, the determination of the structure factor is often difficult and requires, for example, contrast variation experiments.[29] While for hard spheres, the form factor will not be significantly affected by concentration, microgels can deform, interpenetrate and even change their size in dependence of concentration. This hampers the simple calculation of the structure factor by $S(q) = I_{conc.}(q) / I_{dil.}(q)$ with the measured scattering profiles of the dense sample ($I_{conc.}(q)$) and a dilute reference sample ($I_{dil.}(q)$) where $S(q) \approx 1$.[26] In our case, we deal with CS microgels that have a high contrast AuNP core. Due to the much greater volume of the PNIPAM shell with respect to the core, interparticle interactions of our swollen CS microgels are governed by the soft repulsive potential of the shells. Only at extremely high packing densities a contribution from the hard core might become relevant. In this work we focus on packing fractions where the phase behaviour is comparable to purely organic PNIPAM microgels without hard cores similar to our previous work.[40] Specifically, we look at a range of concentrations where we observe the self-organization into crystalline structures.[21, 22] Since the interparticle spacing, $d$, is much larger than the AuNP core size in these samples, we expect the dominating form factor of the cores and the structure factor of the crystalline assemblies to appear in significantly different ranges of $q$.

The as-prepared concentration range of CS particles we investigated spanned from 0.5 wt% to 22.5 wt%. Already visual inspection of these samples provides first insights into the phase behaviour. For the two most dilute concentrations with 0.5 wt% ($\phi_{cs} = 0.06$) and 2 wt% ($\phi_{cs} = 0.24$), fluid-like samples were obtained indicative by the absence of opalescence. In contrast, for concentrations higher than 2 wt% opalescence was observed indicating the formation of crystallites.[14, 61, 62] We also ascribe potential crystallization at lower volume fractions than expected ($\phi < 0.49$) to electrostatic interactions between the slightly charged CS microgels. **Figure 3a** shows scattering profiles from synchrotron SAXS measurements of all investigated samples. Like in the dilute concentration regime investigated with in-house SAXS, the mid to high $q$-regions (> 0.1 nm$^{-1}$) are dominated by the form factor contribution of the AuNP cores. The continuous increase in $I(q)$ at a given $q$ in this region reflects the increase in number density $N$ (see equation 1). Using the forward scattering intensity $I_0$ of the AuNP core scattering determined from the dilute sample and applying a correction that considers the residual water of the freeze-dried CS microgels we can use this dilute concentration of 0.5 wt% to precisely map $N$ for each sample in the concentrated regime. In other words, we can determine the concentrations in the dense phases non-destructively and also account for local density fluctuations and potential sample inhomogeneities. **Table S5** in the Supporting Information lists the as-prepared and corrected wt% concentrations and corresponding volume fractions for all samples investigated. With our set of samples in the concentrated regime we cover a broad range of volume fractions ranging from $\phi_{cs}$ = 0.24 to $\phi_{cs}$ = 1.95. We want to note that values well above the hard sphere packing limit are possible because of the soft and deformable nature of the CS microgels.

Now turning to the low $q$-region in our synchrotron SAXS data, we observe clear structure factor contributions for all concentrated samples with $\phi_{cs} \geq 0.24$. Although the dilute sample (light brown circles in **Figure 3a**) does show an intensity increase with decreasing $q$ in the low $q$-region related to the PNIPAM shells (**Figure S9** in the Supporting Information), the low $q$ scattering of the dense samples is dominated by the structure factor that results from the periodic arrangement of the AuNP cores. In particular for volume fractions close to and higher than the hard sphere packing limit ($\phi$ = 0.74), we expect any form factor contributions from the shell to be negligible. In addition, since the shells will be in close contact and can deform at high $\phi_{cs}$ leading to a reduction in shell contrast.[32, 59] To confirm this assumption, we simulated scattering profiles assuming various shell contrasts and fixed core contrasts. From these simulated profiles we extracted the structure factors shown in **Figure S10** in the Supporting Information. No significant changes in the position of the first structure factor maximum, $q_{max}$, could be detected. A similar effect of disappearing contrast between shell and background can be seen in the UV-Vis spectra from such colloidal crystals where the scattering at lower wavelength is strongly supressed.[40] Assuming that the form factor and structure factor are well separated, we can now eliminate the form factor contribution by dividing all measured intensity profiles by the form factor of the AuNP cores, that is $I_{dil.}(q)$. **Figure 3b** shows the corresponding results for the sample with $\phi_{cs}$ = 0.66. The black circles represent the experimental scattering profile and the red dashed line corresponds to the form factor fit. The extracted structure factor is represented by the blue circles. For $q > 0.08$ nm$^{-1}$ the obtained structure factor is very close to unity highlighting that the form factor contribution was completely eliminated.

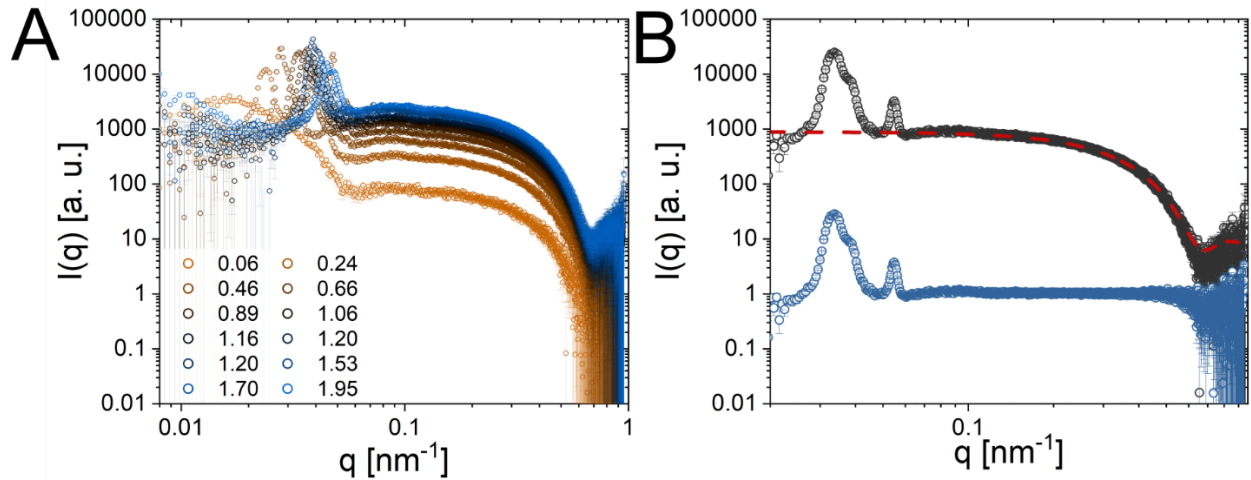

**Figure 3.** A) Radially averaged synchrotron SAXS profiles for a broad range of CS microgel volume fractions listed in the legend. B) Extracted structure factor of the sample with $\phi_{cs}$ =0.66 (blue circles) from the radially averaged intensity profile (black circles) via the fitted AuNP core form factor (red dashed line).

We now want to investigate the structure factors of the samples in more detail. To determine which crystal lattice was formed by the CS microgels, we first analysed the 2D SAXS patterns to determine the origin of the $S(q)$ peaks. As shown in the Supporting Information the 2D detector images of two selected samples reveal at least two orders of pronounced Bragg peaks with a six-fold symmetry indicating the formation of hexagonal close packed planes aligned to the capillary wall (see **Figure S15a** and **S15b** in the Supporting Information). Furthermore, the first order Bragg peaks are superimposed with a pronounced amorphous ring indicating a liquid-like contribution. Thus, these selected samples show the coexistence of crystalline and fluid-like phases in the probed scattering volume. For colloidal spheres it is well known that the stacking sequences of the close-packed hexagonal planes can lead to the formation of a random hexagonally close packed (rhcp) structure, which is a mixture of hexagonally close packed (hcp) and face centred cubic (fcc) stacking sequences, as the energetic differences are very small.[21, 63, 64] To get an idea of the dominant stacking sequence and lattice spacing we modelled the 2D SAXS patterns with the software Scatter (see **Figure S15c** and **S15d** in the Supporting Information). We found a good agreement between the experimental and simulated patterns with a dominant fcc structure consisting of CS microgels with a homogeneous density shell and small domain sizes[65] (for more details see Supporting Information). This finding is in good agreement with the structure found for quite similar CS microgels in a previous SANS study.[40] We want to note that in other works on PNIPAM microgels rhcp phases were identified and that the presence of higher order Bragg peaks is needed to conduct a precise determination of the exact crystal system.[63, 66] The observation of sharp Bragg peaks indicates that the AuNP cores are well centred in the PNIPAM shells, since positional fluctuations would lead to the smearing or absence of clear structure factor peaks.

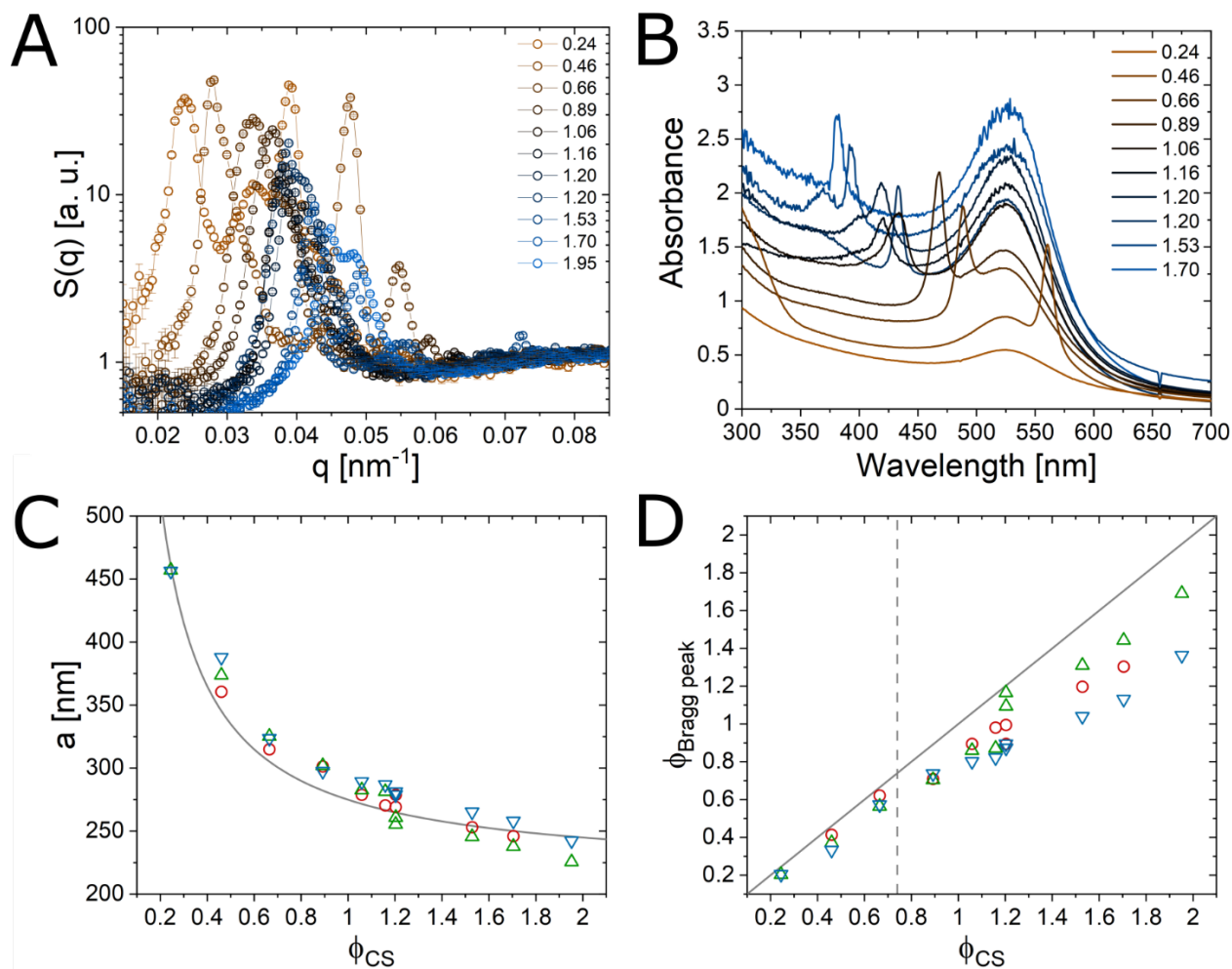

**Figure 4.** A) Extracted structure factors for all crystalline samples. B) UV-Vis absorbance spectra of all crystalline samples recorded in transmission geometry. Volume fractions regarding are listed in the legends. C) Lattice constant ***a*** and (D) volume fraction $\phi_{\text{Bragg peak}}$ extracted from SAXS via the first (111) fcc (blue triangle) and second (220) fcc structure factor peak (green triangle), as well as from the diffraction peaks ($\lambda_{\text{diff}}$) in UV-Vis spectra (red circles). The black curve in (C) serves as guide to the eye and follows the expected scaling between ***a*** and $\phi_{\text{CS}}$. The dashed, vertical line in (D) represents the hard sphere limit for close packed spheres, and the black, solid line corresponds to a linear dependence with a slope equals to one and a zero intercept. Error bars in (C, D) are the same size or smaller than the data points.

**Phase Behaviour.** We now want to focus on the crystalline structures and extract their lattice constants from the Bragg peaks obtained via UV-Vis spectroscopy and from the structure factor peaks determined by SAXS to get insights in the unit cell dimensions of the crystals. From the lattice constants we can also determine the volume fraction of CS microgels.

**Figure 4a** shows the experimentally determined structure factors from SAXS for samples of a broad range of volume fractions, i.e. $\phi_{cs}$ = 0.24 to $\phi_{cs}$ = 1.95. Having verified the alignment of the hexagonally packed planes to the capillary wall and a dominant fcc crystal structure, we can index the peaks in the structure factor profiles with their Miller indices $h$, $k$ and $l$.[26] We can assign the first peak of $S(q)$ to the quasi-forbidden (111) reflection for fcc[65] and the second peak to the (220) reflection. We do note that the first Bragg peaks could also be related to the presence of rhcp stacking that would lead to the appearance of the hcp (100) reflection[67, 68], but its position in $q$ is very close to the (111) peak and therefore the two peaks are hard to distinguish. For simplicity and in agreement to our previous results from SANS[40] and works by others on microgel crystallization, we will continue with the fcc analysis.[26, 69] Here we also want to mention the work of Lapkin *et al.* where a similar system was investigated with larger focus on the crystal structures.[21] In **Figure 4a**, upon increasing $\phi_{cs}$ all peaks move to higher $q$ values indicating a decrease in lattice constant. In addition, the intensity of the structure factor peaks decreases for increasing $\phi_{cs}$ which we attribute to a decrease in the degree of order. In addition, we provide the structure factors that are obtained by dividing using the form factor of the dilute sample in the Supporting Information **Figure S11**. Despite differences in intensity due to the shell contribution the structure factors and in particular the peak positions are in good agreement to the results in **Figure 4a**.

In addition to synchrotron SAXS, we investigated the diffraction properties of our samples using UV-Vis spectroscopy.[22, 40, 70, 71] **Figure 4b** shows the spectra of all samples. The relatively broad peak at $\lambda \approx 524$ nm in all spectra is related to the LSPR of the AuNP cores. In addition, sharp peaks originated from the Bragg reflection of the incident light by the

(111) planes of the crystals can be observed. For instance, for $\phi_{cs}$ = 0.66 the Bragg peak occurs at approximately $\lambda_{Bragg}$ = 488 nm.

Next, we quantitatively compare the results from both techniques, for example via the obtained lattice spacings, $a$. For this, we extracted the first two structure factor peak positions using the Gaussian fits to the data (see **Figure S12** and **S13** in the Supporting Information).

From the structure factor maxima $q_{max}$, we calculate the lattice spacing $d_{hkl}$:

$$d_{hkl} = \frac{2\pi}{q_{max}} \tag{5}$$

The lattice spacing $d_{hkl}$ is assigned to different crystalline planes by the Miller indices $h, k$ and $l$. With the lattice spacing we calculate the lattice constant $a$ for *the* (111) peak according to:

$$a = d_{hkl}\sqrt{h^2 + k^2 + l^2} = d_{hkl}\sqrt{3} \tag{6}$$

The Bragg peak analysis by UV-Vis spectroscopy gives also access to the lattice spacing:

$$m\lambda_{Bragg} = 2d_{hkl}\sqrt{n_{crystal}^2 - \sin\theta^2} \tag{7}$$

Here $\lambda_{Bragg}$ is the spectral position of the diffraction peak, $d_{hkl}$ the lattice spacing, $n_{crystal}$ the refractive index of the crystalline sample and $\theta$ the angle between the incoming plane and the normal to the crystal plane. [33, 52, 62] For the (111) crystalline plane one can simplify the calculation, since we assume the hexagonally close packed planes to be oriented parallel to the capillary wall so that $\theta$ = 0°:

$$d_{111} = \frac{\lambda_{diff}}{2n_{crystal}} \tag{8}$$

We use $n_{crystal}$ = 1.345 as the average refractive index of the colloidal crystals similar to our previous work.[40]

**Figure 4c** compares the determined lattice constants from SAXS and UV-Vis spectroscopy in dependence of the volume fraction of the samples. The data from both methods nicely collapse onto a single master curve showing a decrease from $a$ = 457 nm at $\phi_{cs}$ = 0.24 to $a$ = 226 nm at $\phi_{cs}$ = 1.95. The black line in **Figure 4c** serves as guide to the eye following the theoretical scaling between $a$ and $\phi_{cs}$, more details are given in the Supporting Information. Here we note that the volume fraction we use does not take into account any interpenetration and/or faceting of microgels when the volume fraction exceeds the hard sphere limit of $\phi$ = 0.74 as recently discussed by Scotti et al.[23, 30]

In agreement to the scaling of $a$ with $\phi_{cs}^{-1/3}$ we found the decrease of $\lambda_{diff}$ to scale with $\phi_{cs}^{-1/3}$ which was previously observed for microgel systems (see **Figure S14** in the Supporting Information).[15]

We now want to compare the relation between the volume fraction $\phi_{cs}$ that we determined using the forward scattering of the AuNP cores in SAXS and the volume fraction $\phi_{Bragg\ peak}$ that we can calculate from the Bragg peak analysis of SAXS and UV-Vis data. Using the lattice constants, $a$ and the hydrodynamic radii, $R_h$ we get the following relation:

$$\phi_{Bragg\ peak} = \frac{(3+1)\frac{4}{3}\pi R_h^{\ 3}}{a^3} \tag{9}$$

Here an fcc lattice was considered and the (3 + 1) terms corresponds to the number of CS microgels in the fcc unit cell. In **Figure 4d** we compare $\phi_{Bragg\ peak}$ obtained from SAXS and UV-Vis data with $\phi_{cs}$. The data from both techniques are in good agreement and we find a linear scaling between $\phi_{Bragg\ peak}$ and $\phi_{cs}$ for low volume fractions. At higher volume fractions exceeding the hard sphere limit of $\phi$ = 0.74 (dashed line in **Figure 4d**), we find that $\phi_{Bragg\ peak}$ is significantly smaller than $\phi_{cs}$. This deviation indicates that microgel deformation and/or interpenetration starts to occur.  In addition, we believe that the formation of wall crystals in coexistence with a disordered sample structure, as also indicated by the presence of the pronounced rings in the 2D SAXS patterns (**Figure S15** in the Supporting Information) can cause this discrepancy. The discrepancy between $\phi_{Bragg\ peak}$ determined from the (111) and the (220) reflections for samples exceeding the apparent volume fraction of 1 might be related to contributions from rhcp lattices as mentioned before. We can also not exclude that the form factor of the CS microgels in the dense regime might influence the peak positions to some extent.

Clearly, the phase behaviour and packing in dense microgel systems is difficult to quantify precisely. The benefits from scattering techniques for the investigation of soft colloids like microgels at higher volume fractions are clearly the good statistics by probing millions of particles at once. SANS is often used to investigate such samples but suffers from long durations for a single measurement.  The combination of a hard inorganic core encapsulated in a soft and responsive microgel shell provides us with a powerful model system to make use of SAXS with acquisition times below 1 s and the separation between the form factor and structure factor, due to the difference in size and contrast between core and shell. When normalized to absolute scattering intensities, the scattering signal of the core also provides detailed information about the number concentration of particles, which makes this system promising not only for dense packing, but also for studies in the dilute state were accurate control over the particle number is needed.

## Conclusion

We have studied core-shell microgels in the dilute and densely packed regime using small-angle X-ray scattering. Owed to the pronounced difference in volume of the gold nanoparticle cores and the cross-linked poly-$N$-isopropylacrylamide shells as well as the stark difference in the scattering contrast, form factor and structure factor of crystalline samples are well separated. Using absolute scattering intensities and form factor analysis of the gold core contribution, we could determine microgel number concentrations very precisely. With this we could relate the experimentally given weight concentration to microgel volume fractions. We find that the microgels form crystals with a dominant fcc structure over a large range of $\phi_{cs}$. The quantitative agreement between crystal analysis from SAXS and results from UV-Vis spectroscopy is very good, with the lattice constants of the crystal unit cells collapsing on a single master curve. There is an increasing deviation for higher volume fractions, in particular, above the hard sphere packing limit between the volume fraction obtained from the number concentration of the cores and the volume fraction determined from the lattice constant. This can be explained by having crystals in co-existence with a disordered sample structure indicated by distinct rings in the SAXS patterns and the occurrence of microgel deformation and interpenetration.

We propose that such core-shell microgels are ideal model systems to study the phase behaviour of microgels, in particular at high volume fractions. While the small gold nanoparticle cores dominate the mid to high $q$ scattering in SAXS, they do not significantly affect the inter-particle interactions that are governed by the microgel shells. The short acquisition times of SAXS open up the possibility to investigate the impact of temperature changes on the microgels in dense packing, which induces melting and recrystallisation of the system, due to the temperature-dependent volume of the microgels[21], but has also been shown to change the particle interactions.[72] In addition, core-shell microgels with high contrast cores will allow further explorations of other fundamental phenomena such as jamming and the glass transition of soft colloidal systems. This is expected to lead to new insights in the field of soft colloidal systems.

## AUTHOR INFORMATION

### Corresponding Authors

\* Prof. Dr. Matthias Karg

Institut für Physikalische Chemie I: Kolloide und Nanooptik,

Heinrich-Heine-Universität Düsseldorf, Universitätsstraße 1, D-40225 Düsseldorf, Germany

E-Mail: karg@hhu.de

\*Ass. Prof. Dr. Janne-Mieke Meijer

Department of Applied Physics and Institute for Complex Molecular Systems, Eindhoven University of Technology, P.O. Box 513, 5600 MB Eindhoven, The Netherlands

E-Mail: j.m.meijer@tue.nl

### Funding Sources

J.M.M. acknowledges financial support from the Netherlands Organization for Scientific Research (NWO) (016.Veni.192.119). M.K. acknowledges the German Research Foundation (DFG) for funding under grant KA3880/6-1.

### Notes

The authors declare no competing financial interest.

### Supporting Information

Additional TEM images and analysis; UV-Vis spectra of dilute AuNP core and CS microgel dispersions; temperature dependent DLS measurements; form factor and Guinier analysis; determination of number density and volume fraction; structure factor analysis; characterization of crystalline samples.

## ACKNOWLEDGMENTS

We acknowledge SOLEIL for provision of synchrotron radiation facilities and we would like to thank Javier Perez for assistance in using beamline SWING (Proposal number 20181613). The authors thank the Center for Structural Studies (CSS) that is funded by the Deutsche Forschungsgemeinschaft (DFG Grant numbers 417919780 and INST

208/761-1 FUGG) for access to the SAXS instrument. The authors acknowledge the DFG and the state of NRW for funding the cryo-TEM (INST 208/749-1 FUGG) and Marius Otten from Heinrich-Heine-University Düsseldorf for his assistance with the TEM measurements.

Insert Table of Contents artwork here

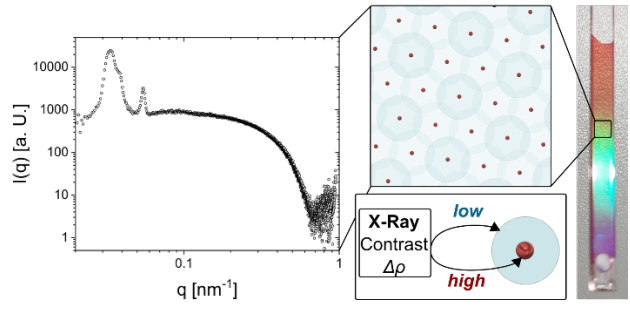

# Supporting Information

# SAXS Investigation of Core-Shell Microgels with High Scattering Contrast Cores: Access to Structure Factor and Volume Fraction

Marco Hildebrandt,[1] Sergey Lazarev,[2] Javier Pérez,[3] Ivan A. Vartanyants,[2] Janne-Mieke Meijer, [4,*] Matthias Karg[1,*]

[1]Institut für Physikalische Chemie I: Kolloide und Nanooptik, Heinrich-Heine-Universität Düsseldorf, Universitätsstraße 1, D-40225 Düsseldorf, Germany

E-Mail: karg@hhu.de

[2]Deutsches Elektronen-Synchrotron DESY, Notkestraße 85, 22607 Hamburg, Germany

[3]Synchrotron SOLEIL, L'Orme des Merisiers, Saint-Aubin, BP 48, 91192 Gif-sur-Yvette Cedex, France

[4]Department of Applied Physics and Institute for Complex Molecular Systems, Eindhoven University of Technology, P.O. Box 513, 5600 MB Eindhoven, The Netherlands

E-Mail: j.m.meijer@tue.nl

**General Characterization of Core-Shell Microgels**

The morphology of the CS microgels and the successful encapsulation of single AuNP cores in each microgel was investigated by TEM. **Figure S1** shows TEM images of the CS microgels at different magnifications. These images were used to determine the encapsulation rate of the AuNP cores in the PNIPAM shells.

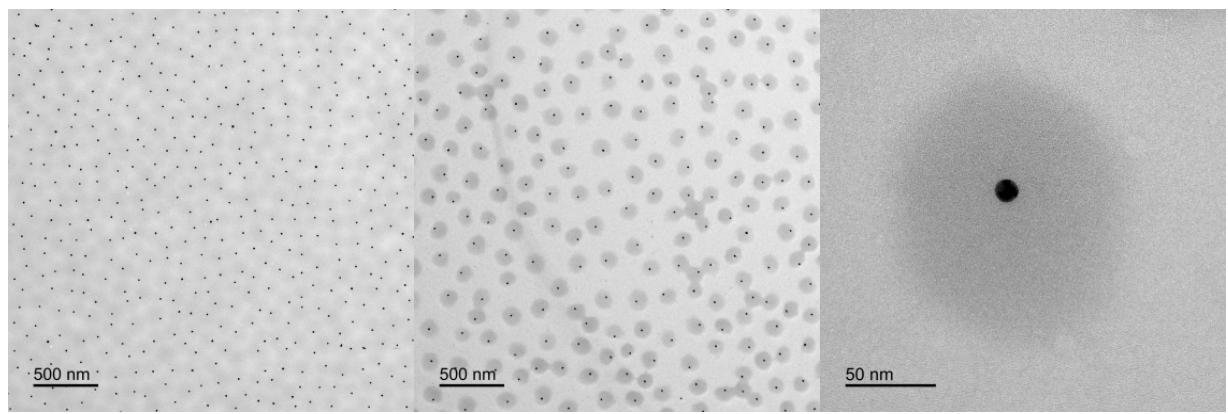

Figure S1. TEM images of CS microgels recorded at different magnifications.

**Figure S2** shows the size distribution histogram of the AuNP cores as obtained from manual TEM image analysis using ImageJ.

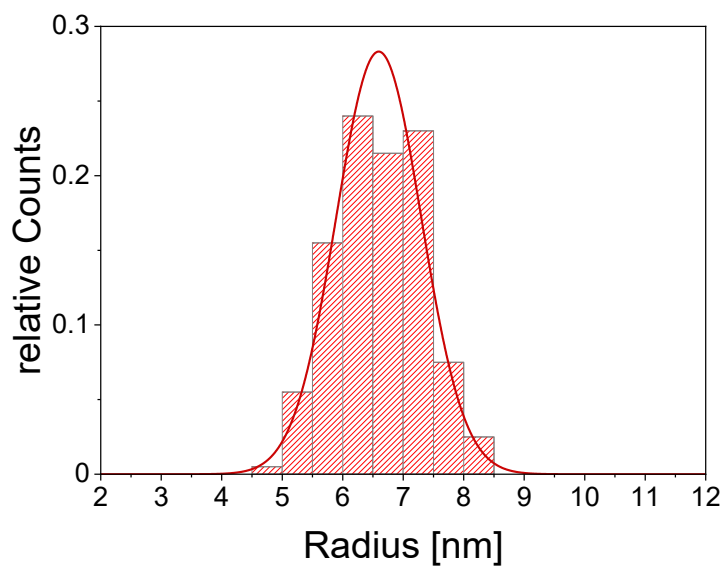

**Figure S2.** Size distribution of the AuNP cores as determined by TEM. The red line corresponds to a fit to the data using a Gaussian distribution function.

The optical properties of the AuNP cores and the final CS microgels were studied by UV-Vis spectroscopy. **Figure S3** shows UV-Vis absorbance spectra of the as-synthesized AuNP cores and the final Au-PNIPAM CS microgels. The $Au^0$ concentration of the AuNP core dispersion was determined via the absorbance at 400 nm, $Abs(400)$, according to:[1]

$$c_{Au^0} = \frac{Abs(400) \times F_D}{\varepsilon} \qquad \textbf{(S.1)}$$

Here $\varepsilon$ = 2330 Lmol⁻¹cm⁻¹ and $F_D$ is the factor of dilution.

The peaks in absorbance at λ = 524 nm (vertical red lines) are related to the localized surface plasmon resonance (LSPR) of the AuNPs. For the CS microgels the LSPR is less pronounced due to the increased light scattering contribution of the polymer shell (**Figure S3B**).

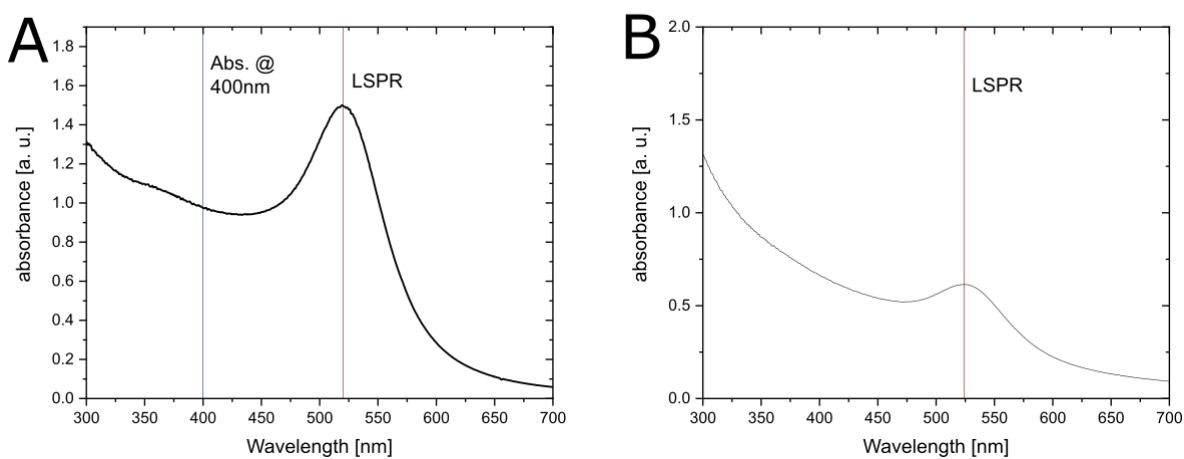

**Figure S3.** A) Absorbance spectrum of the AuNP core dispersion ($F_D$ = 50). B) Absorbance spectrum of the CS microgels in dilute aqueous dispersion. The vertical red lines highlight the LSPR positions of the AuNPs.

The volume phase transition (VPT) of the CS microgels was followed by temperature-dependent DLS measurements. **Figure S4** shows the resulting evolution of the hydrodynamic radius, $R_h$, as a function of temperature. $R_h$ decreases continuously with increasing temperature until reaching nearly constant values at temperatures of 60 °C and higher where the PNIPAM shells are in their collapsed state. The VPT temperature (VPTT) is approximately 36.2 °C. Even at high nominal cross-linker contents of 15 mol% and more PNIPAM microgels show distinct response to temperature.[2, 3]

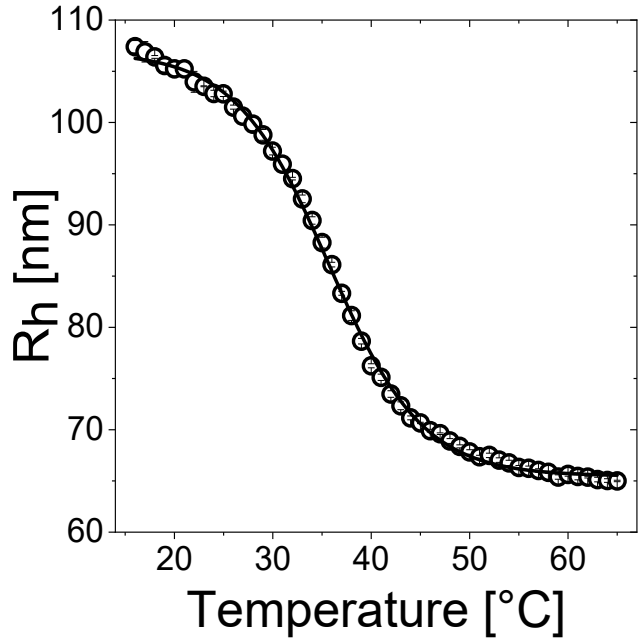

**Figure S4.** Hydrodynamic radii ($R_h$) of the CS microgels from DLS as a function of temperature. The solid black line corresponds to a fit using the Boltzmann sigmoidal function.

## Form Factor Modeling

We measured the form factor of the CS microgels in dilute dispersion using SAXS. Due to the isotropic character of all recorded scattering data in the dilute regime, data were radially averaged to obtain the average scattering intensity, $I$, as a function of the magnitude of the scattering vector, $q$, given by:

$$|\vec{q}| = q = \frac{4\pi}{\lambda} sin\left(\frac{\theta}{2}\right) \tag{S.2}$$

Here $\theta$ is the scattering angle and $\lambda$ the wavelength of the X-ray beam.

For the investigated $q$-range the scattered intensity, $I(q)$, depends on the microgel number density, $N$, the volume of a single microgel, $V_p$, the scattering length density (SLD) difference between the scattering object and the solvent, $\Delta SLD$, the form factor $P(q)$ of the scattering object and additional background contributions, $I_B$:

$$I(q) = NV_P^2(\Delta SLD)^2 P(q) + I_B \tag{S.3}$$

Here we neglect any structure factor contributions, $S(q)$, since we are in the dilute regime, where $S(q) \approx 1$.

The form factor of a solid sphere with the radius $R$ is given by:

$$P(q) = \left[3\frac{sin(qR)-qRcos(qR)}{(qR)^3}\right]^2 \tag{S.4}$$

The form factor of a spherical shell is given by the radius of the core $R$, the thickness of the shell $\Delta R$ and the difference in scattering length dentistry between the matrix and the respective core and shell, $\Delta SLD_{core}$ and $\Delta SLD_{shell}$.

$$I_{shell}(q, R, \Delta R, \Delta SLD_{core}, \Delta SLD_{shell})$$
$$= [K(q, R + \Delta R, \Delta SLD_{shell}) - K(q, R, \Delta SLD_{shell} - \Delta SLD_{core})]^2 \tag{S.5}$$

with

$$K(q, R, \Delta SLD) = \frac{4}{3}\pi R^3 \Delta SLD \, 3\frac{sin(qR)-qRcos(qR)}{(qR)^3} \tag{S.6}$$

In order to account for polydispersity, the form factor and the volume are convoluted with a normalized Gaussian distribution function:

$$D(R, \langle R \rangle, \sigma_{poly}) = \frac{1}{\sqrt{2\pi\sigma_{poly}^2}}exp\left(-\frac{(R-\langle R \rangle)^2}{2\sigma_{poly}^2}\right) \tag{S.7}$$

Here $\langle R \rangle$ is the average particle radius and $\sigma_{poly}$ is the relative size polydispersity. In order to describe the measured form factor of the CS microgels, we used the simple sum of two homogeneous spheres. Thus, one polydisperse sphere accounts for the scattering of the AuNP cores, while the second one accounts for scattering from the PNIPAM shell. We want to highlight that this simple superposition does not consider for any interference between core and shell. Due to the large difference in contrast and size between the AuNP cores and the PNIPAM shells as well as the limited resolution of the microgel form factor, we used this very simplified model to estimate the total microgel size from our SAXS data. Since we do not attempt to describe the microgel form factor in detail in this work, we do not apply more complex core-shell or core-shell-shell models that are commonly used for microgels/core-shell microgels.[4-6] **Table S1** lists the obtained fit parameters for the core and the shell contribution obtained from a measurement of a diluted (0.5 wt%) sample at 20°C.

**Table S1.** Summary of fit parameters obtained from form factor analysis using SASfit.

|  | **Core** | **Shell** |
|---|---|---|
| $R$ | 6.5 ± 0.6 nm | 77.9 ± 8.7 nm |
| $\Delta SLD$ | $1.2 \times 10^{-2}$ nm$^{-2}$ | $4.7 \times 10^{-5}$ nm$^{-2}$ |
| $I_B$ | 0.35 cm$^{-1}$ | 0.35 cm$^{-1}$ |
| $\sigma_{poly}$ | 0.09 | 0.11 |

In addition to the sum of the two polydisperse spheres (red line), we show form factor fits based on a core-shell model in **Figure S5**. The two fits based on the core-shell model only differ in the polydispersity ($\sigma_{poly} = 0.1$ for the fit in green and $\sigma_{poly} = 0.24$ for the fit in blue). More details on the fits are listed in **Table S2**. We see that the fit exhibiting the higher polydispersity describes the data quite well but the unrealistically high polydispersity of $\sigma_{poly} = 0.24$ is in strong contrast to former SANS studies on very similar systems resulting in polydispersities close to $\sigma_{poly} = 0.1$. Here we want to note that the contrast of the polymer shell is much higher in SANS and should lead to more reliable results.[2, 3]

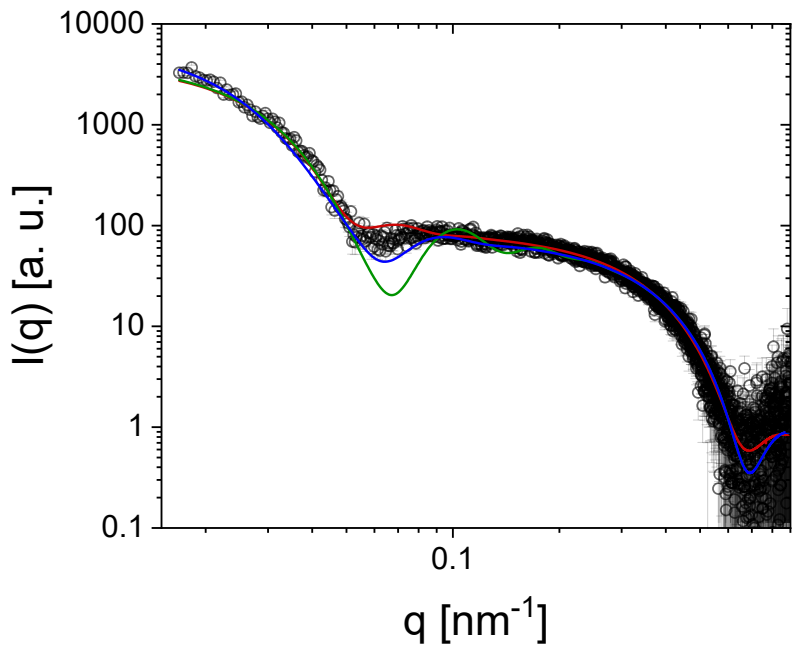

**Figure S5.** SAXS profile from dilute CS microgel dispersion (20°C). The solid lines correspond to the applied form factor fits. In red, the introduced fit based on the sum of two polydisperse spheres, in blue and green fits based on the core-shell model with polydispersities of $\sigma_{poly} = 0.1$ (green) and $\sigma_{poly} = 0.24$ (blue) of the respective shell.

**Table S2.** Summary of fit parameters obtained from form factor analysis based on the core-shell model using SASfit.

|  | **Core-Shell** | **Core-Shell** |
|---|---|---|
| $R_{core}$ | 6.5 nm | 6.5 nm |
| $R_{shell}$ | 79.6 ± 8 nm | 79.6 ± 19.2 nm |
| $I_B$ | 0.35 cm$^{-1}$ | 0.35 cm$^{-1}$ |
| $\Delta SLD_{core}$ | 1.2 × 10$^{-2}$ nm$^{-2}$ | 1.2 × 10$^{-2}$ nm$^{-2}$ |
| $\Delta SLD_{shell}$ | 3.2 × 10$^{-5}$ nm$^{-2}$ | 3.2 × 10$^{-5}$ nm$^{-2}$ |
| $\sigma_{poly, shell}$ | 0.1 | 0.24 |

The radius of gyration $R_g$ of the CS microgels was determined by analysis of the Guinier region at low $q$ as shown in **Figure S6**.

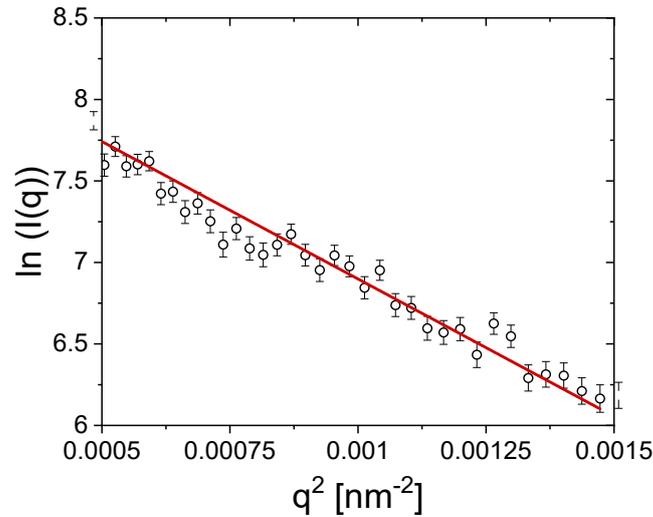

**Figure S6.** Guinier plot and linear fit (red line) of the low $q$ SAXS data measured form a dilute CS microgel dispersion (0.5 wt%) at 20 °C.

The slope of the linear fit gives access to $R_g$:

$$I(q) = I_0 \times exp\left[\frac{-q^2 R_g^2}{3}\right] \tag{S.8}$$

We obtain $R_g$ = 71.1 ± 1.7 nm.

Due to the large difference in core and shell size, our SAXS data from dilute samples (see **Figure 2a** of the main manuscript) allow also to perform Guinier analysis in the mid to high $q$-range, where the scattering of the AuNP cores dominates. Thus, we can determine the values of $R_g$ and also the forward scattering intensity, $I_0$. **Figure S7** shows the corresponding Guinier plots and linear fits to the data for three different CS microgel concentrations.

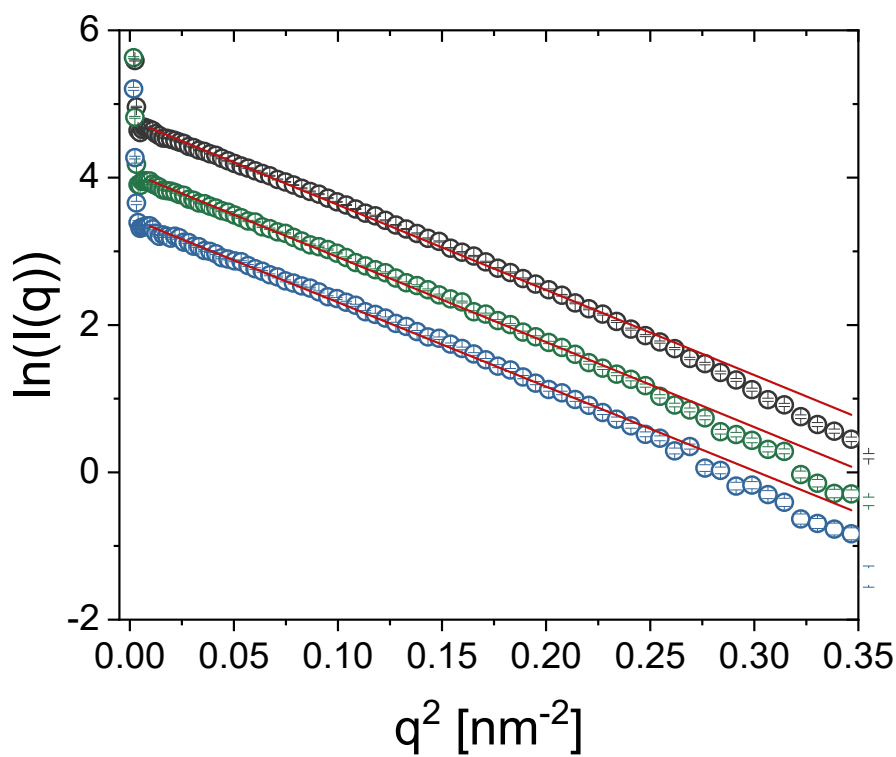

**Figure S7.** Guinier plots of the SAXS data measured from dilute CS microgel dispersions with 2.45 wt% (black circles), 1.23 wt% (green circles) and 0.62 wt% (blue circles). Linear fits (red lines) were applied to extract $I_0$ and $R_g$ according to equation S.6.

The parameters from the linear fit and the resulting forward scattering $I_0$ and $R_g$ are listed in **Table S3**.

**Table S3.** Parameters from Guinier analysis of the AuNP core scattering contribution at mid to high $q$.

|  | **2.45 wt%** | **1.23 wt%** | **0.62 wt%** |
|---|---|---|---|
| Intercept | 4.79 | 4.07 | 3.45 |
| Slope [nm$^{-2}$] | -11.55 | -11.53 | -11.44 |
| $I_0$ [cm$^{-1}$] | 120 | 59 | 32 |
| $R_g$ [nm] | 5.89 | 5.88 | 5.86 |

As expected, the forward scattering scales with the microgel concentration. The obtained values of $R_g$ are slightly smaller than the value obtained from the polydisperse sphere form factor analysis ($R_{core}$ = 6.54 nm). This is expected for homogeneous spheres.

## Determination of Number Density and Volume Fraction

With our absolute intensity in-house SAXS data and the theoretical scattering length densities of the AuNP cores and water, we can calculate the particle number density, $N$, based on the intensity $I_0$ at infinitely small $q$, i.e. the forward scattering intensity of the AuNP cores:

$$N = \frac{I_0 \, N_A \, \rho^2}{m \, M_w \, \Delta SLD^2}$$  **(S.9)**

Here $N_A$ is Avogadro's number, $\rho$ the density of the AuNP cores (19.3 g·cm$^{-3}$), $m$ the average mass of a single AuNP core and $M_w$ its molecular weight. With the values of $I_0$ obtained from the Guinier analysis provided in **Figure S7**, we calculated $N$ (**Table S4**).

Table S4. Number concentrations obtained from SAXS measurements of diluted samples with known concentrations.

|  | **2.45 wt%** | **1.23 wt%** | **0.62 wt%** |
|---|---|---|---|
| N [10$^{13}$ mL$^{-1}$] | 6.59 | 3.22 | 1.73 |

Since we know from detailed TEM analysis that each CS microgel contains one single AuNP core, the obtained number density of the AuNP cores is equal to the number density of CS microgels, i.e. the number of cores is equal to the number CS microgels. This allows to determine the volume fraction of CS microgels, $\phi_{cs}$:

$$\phi_{CS} = N \, \frac{4}{3} \pi \, R_h{}^3$$  **(S.10)**

**Figure S8** shows the determined values of $\phi_{cs}$ as a function of the weight concentration (wt%). The linear fit to the data (black line) for the differently concentrated samples in the dilute regime allow for an extrapolation to higher concentrations. We find the following relation:

$$\phi_{CS} = (0.130 \, \pm \, 0.001) \, \frac{\text{mass concentration in wt\%}}{\text{wt\%}}$$  **(S.11)**

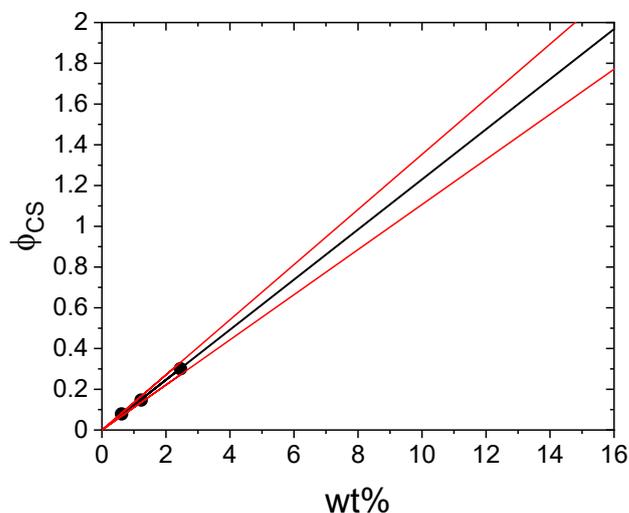

**Figure S8.** Volume fraction of CS microgels $\phi_{CS}$ determined for samples in the dilute regime from in-house SAXS measurements. The black line is a linear fit to the data. The red lines indicate a 10% error corridor.

The very valuable relation provided by equation **S.11** and the fact that we can precisely determine the forward scattering of the AuNP core contribution allows us to determine the local concentration of dense samples probed by the rather small X-ray beam in the synchrotron SAXS experiments. As reference for the mapping of the concentrations, we choose the most dilute sample (0.5 wt%). This sample was prepared by dilution of a stock dispersion prepared from freeze-dried CS microgels. We considered a residual water content in the freeze-dried sample of 5.7%.[2] **Table S5** lists the results for all sample concentrations studied in this work. The as-prepared, expected concentrations, are the concentrations that were aimed at by weighing in freeze-dried microgels and dispersing the microgels in the respective amounts of water. The forward scattering intensities from the AuNP core scattering contribution, $I_0$, were determined by Guinier analysis of the mid to high-$q$ scattering region as demonstrated in **Figure S7**.

**Table S5.** Determination of sample concentration and volume fraction based on the forward scattering of the AuNP core contribution in the synchrotron SAXS measurement and the relation given by equation **S.11**.

| as-prepared expected concentration [wt%] | $I_0$ [a.u.] | concentration based on $I_0$ [wt%] | concentration[a] corrected for water content [wt%] | volume fraction $\phi_{cs}$ |
|---|---|---|---|---|
| 0.5 | 88.6 | 0.5 | 0.47 | 0.06 |
| 2 | 353 | 1.99 | 1.88 | 0.24 |
| 4 | 665 | 3.75 | 3.54 | 0.46 |
| 6 | 961 | 5.42 | 5.11 | 0.66 |
| 8 | 1290 | 7.28 | 6.86 | 0.89 |
| 10 | 1530 | 8.63 | 8.14 | 1.06 |
| 11 | 1675 | 9.45 | 8.91 | 1.16 |
| 12.5 | 1739 | 9.81 | 9.25 | 1.20 |
| 11 | 1740 | 9.82 | 9.26 | 1.20 |
| 15 | 2210 | 12.47 | 11.76 | 1.53 |
| 17.5 | 2464 | 13.9 | 13.11 | 1.70 |
| 22.5 | 2822 | 15.92 | 15.02 | 1.95 |

[a]Values were corrected for a residual water content of 5.7% (by mass).

## Structure Factor Analysis

We now turn to the synchrotron SAXS investigation of dense samples in the crystalline regime. We first want to address the potential influence of the scattering contribution from the CS microgels at low $q$ as observed from the synchrotron SAXS measurements in the dilute regime. To do so we divide the experimental scattering profile of the dilute sample by the fitted polydisperse sphere form factor of the AuNP cores $P_{core}(q)$. The residual scattering profile shown in **Figure S9** should correspond solely to scattering from the PNIPAM shells. The shell contribution is only visible for $q < 0.07$ nm$^{-1}$. Here we observe a continuous increase in scattering intensity with decreasing $q$.

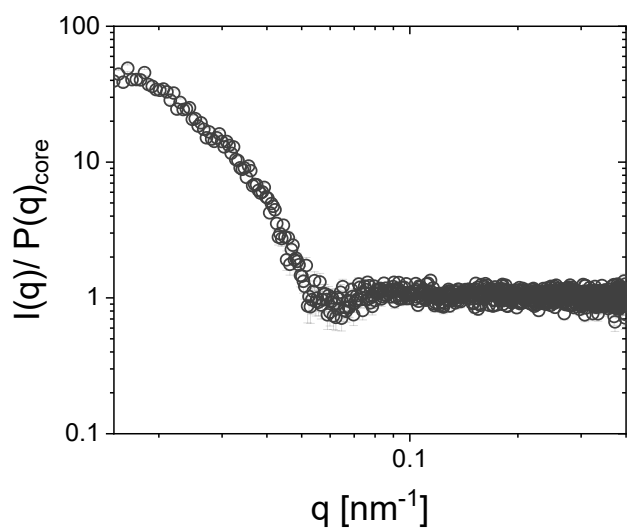

**Figure S9.** Shell contribution to the scattering intensity after division by $P_{core}(q)$ as measured with synchrotron SAXS.

We now want to discuss whether such residual scattering from the PNIPAM shells hampers the analysis of the structure factor with a particular focus on the peak positions of $S(q)$. While for systems of hard, non-deformable scattering objects $S(q)$ is typically easily accessible by simply dividing scattering profiles of dense samples by profiles of dilute samples, where $S(q) \approx 1$, the situation is more complex for soft microgels. In the dense packing regime, microgels can deform and/or interpenetrate and thus, the form factor is different as compared to the dilute state. We want to determine whether this is also the case for our SAXS data of the CS microgels where scattering of the shells is weak and core scattering dominates. To do so, we simulated scattering profiles in SASfit using the form factor and contrast parameters as listed in **Table S1** and a structure factor corresponding to a volume fraction of 0.64 and a hard sphere radius of 115 nm. The specific hard sphere radius and volume fraction are chosen to give the most realistic description of the system and are obtained from the lattice spacing *a*. We used a structure factor based on the Percus-Yevick model for hard sphere fluids.[7] These simulated scattering profiles where then divided by the dilute state form factor only where we apply different contrasts for the PNIPAM shell. A contrast of 1 $\Delta SLD$(shell) corresponds to the values given in **Table S1**. A contrast of 0 $\Delta SLD$(shell) corresponds to scattering of the AuNP cores only. **Figure S10** shows the resulting structure factors for different shell contrasts.

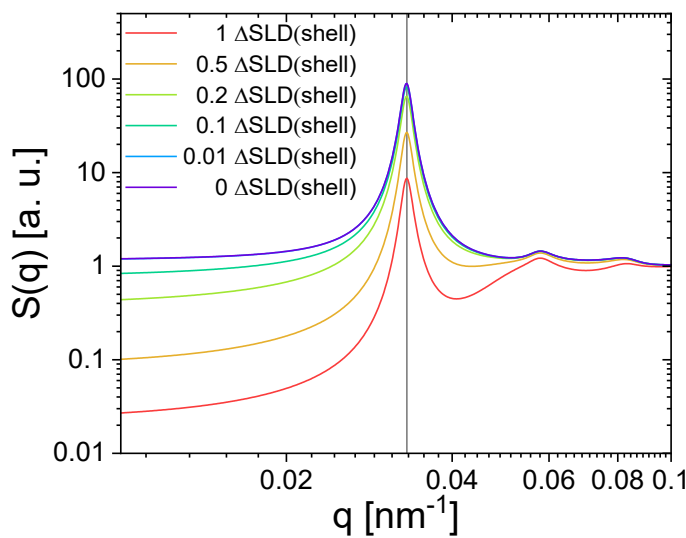

**Figure S10.** Calculated structure factors from SASfit simulations for decreasing shell contrasts. The vertical black line highlights the position of the first structure factor peak.

While we see a pronounced influence in structure factor intensity at low $q$, the position of the structure factor peaks, $q_{max}$, remain unaffected by the shell scattering contribution. The first structure factor peak appears at $q_{max}$ = 0.0331 nm$^{-1}$, independent of the considered shell contrast. Even for the simple division by the AuNP core form factor (0 $\Delta SLD$(shell)), we obtain the same peak positions. Therefore, all structure factors of dense samples shown in the main manuscript were obtained by dividing the experimental SAXS profiles by the AuNP core form factor contribution only.

Here we also want to provide the structure factor extracted by dividing the scattering profile of the dense samples $I_{conc.}(q)$ by the scattering profile of the dilute sample $I_{dil.}(q)$ where $S(q) \approx 1$. **Figure S11** shows the resulting structure factors in the same $q$-range as used in **Figure 4a** in the main manuscript. We see a shift of the structure factor towards high $q$ with increasing volume fraction. Overall, we find very similar peak positions and shapes as in **Figure 4a**. The most prominent differences refer to the peak intensities and overall structure factor intensity at very low $q$ where the shell scattering from the dilute sample is strongest. This is also clearly seen in **Figure S10**.

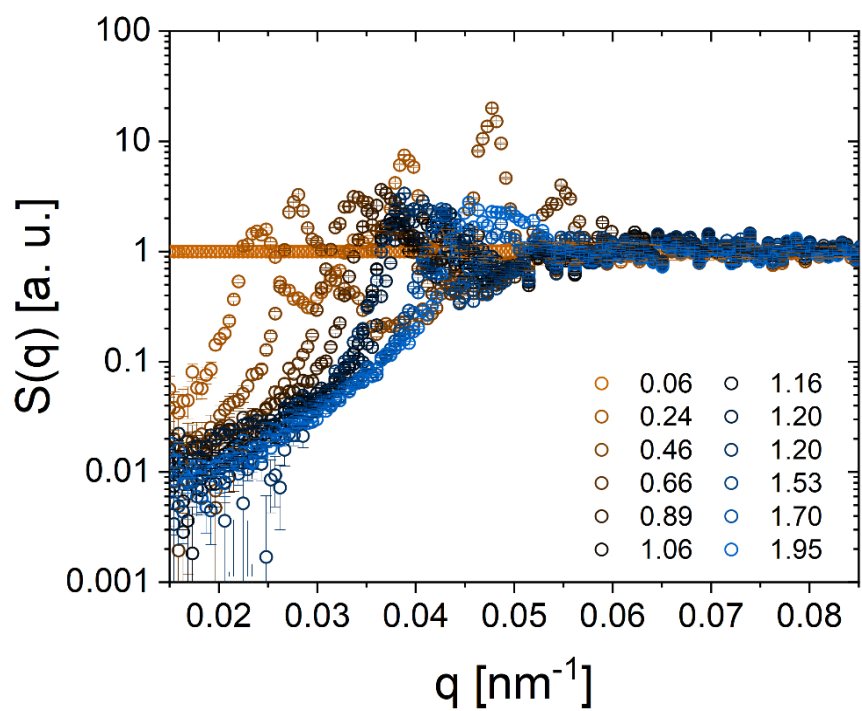

**Figure S11.** $S(q)$ extracted by dividing $I_{conc.}(q)$ through $I_{dil}(q)$ obtained from the experimentally measured SAXS profile of a dilute CS microgel dispersion (20°C).

In **Figure S12** we exemplarily show the obtained structure factor for a CS microgel sample at $\phi_{cs}$ = 0.66.

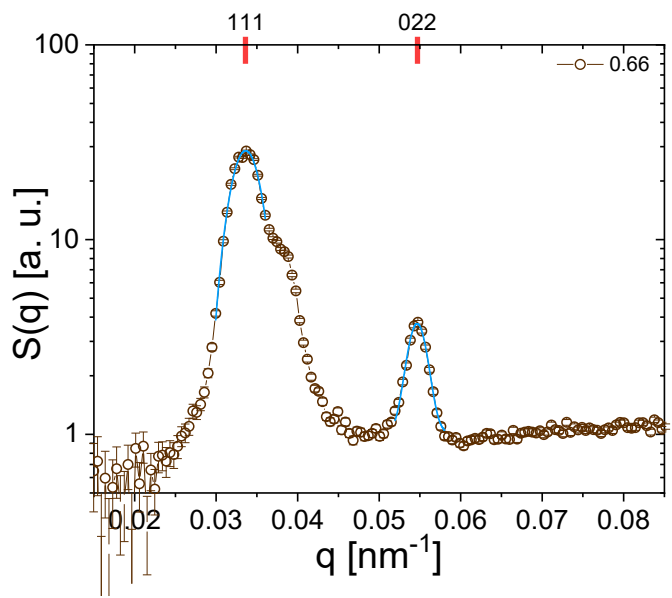

**Figure S12.** S(q) obtained from a sample at a volume fraction of 0.66 and the Peak assignment of the 111 and the 022 peak for a fcc crystals structure including Gaussian fits which were used to extract the structure factor maxima.

The first two structure factor maxima are clearly visible and correspond to the (111) and (022) peaks. The high-$q$ shoulder of the (111) peak corresponds to the amorphous ring from the fluid-like scattering contribution. To extract the positions of the structure factor maxima, we applied Gaussian fits to the data in selected ranges of $q$ (blue lines in **Figure S12**).

For higher volume fractions the intensity of the (022) peaks drops significantly as shown in **Figure S13**. Nevertheless, we could still analyze the peak positions by performing Gaussian fits to the data.

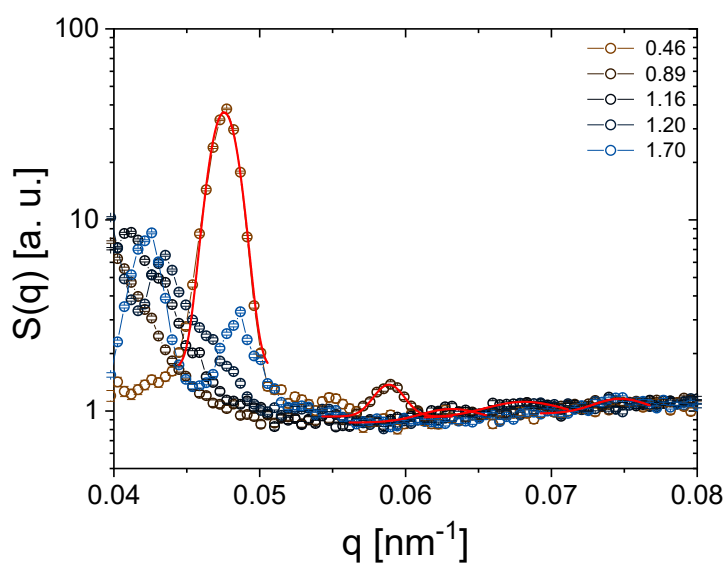

**Figure S13.** Structure factors in a selected $q$-range and analysis of the (022) peaks by Gaussian fits (red lines).

In total we analyzed the structure factor of 11 different samples covering a broad range of volume fractions. In addition to Bragg peak analysis from SAXS, we also studied the sample diffraction by using UV-Vis spectroscopy. The results from both methods are summarized in **Tables S6** and **S7**.

**Table S6.** Positions of the first two structure factor maxima ($q_{max}$), positions of Bragg peaks from UV-Vis spectroscopy ($\lambda_{Bragg}$) and the resulting lattice spacings (d).

| $\phi_{CS}$ | $q_{max}(111)$ [nm$^{-1}$] | $q_{max}(002)$ [nm$^{-1}$] | $\lambda_{Bragg}$ [nm] | $d(111)$ [nm] | $d(022)$ [nm] | $d_{UV\text{-}Vis}(111)$ [nm] |
|---|---|---|---|---|---|---|
| 0.24 | 0.0239 | 0.0389 | | 263 | 162 | |
| 0.46 | 0.0281 | 0.0476 | 560 | 224 | 132 | 208 |
| 0.66 | 0.0336 | 0.0547 | 489 | 187 | 115 | 182 |
| 0.89 | 0.0366 | 0.0589 | 468 | 172 | 107 | 174 |
| 1.06 | 0.0376 | 0.0629 | 433 | 167 | 100 | 161 |
| 1.16 | 0.0379 | 0.0632 | 420 | 166 | 99 | 156 |
| 1.20 | 0.0390 | 0.0696 | 418 | 161 | 90 | 155 |
| 1.20 | 0.0387 | 0.0682 | 433 | 162 | 92 | 161 |
| 1.53 | 0.0410 | 0.0724 | 393 | 153 | 87 | 146 |
| 1.70 | 0.0422 | 0.0747 | 382 | 149 | 84 | 142 |
| 1.95 | 0.0449 | 0.0788 | | 140 | 80 | |

**Table S7.** Lattice constants ($a$) and the resulting volume fractions of the CS microgels calculated from the lattice spacings of **Table S5**.

| $\phi_{CS}$ | $a$ [nm] (111) | $a$ [nm] (022) | $a_{UV\text{-}Vis}$ [nm] | $\phi_{CS}$ (111) | $\phi_{CS}$ (022) | $\phi_{CS}$ UV-Vis |
|---|---|---|---|---|---|---|
| 0.24 | 456 | 457 | | 0.20 | 0.20 | |
| 0.46 | 388 | 374 | 361 | 0.33 | 0.37 | 0.41 |
| 0.66 | 324 | 325 | 315 | 0.57 | 0.56 | 0.62 |
| 0.89 | 298 | 302 | 301 | 0.74 | 0.70 | 0.71 |
| 1.06 | 289 | 283 | 279 | 0.80 | 0.86 | 0.90 |
| 1.16 | 287 | 281 | 270 | 0.82 | 0.87 | 0.98 |
| 1.20 | 279 | 255 | 269 | 0.89 | 1.17 | 0.99 |
| 1.20 | 281 | 261 | 279 | 0.87 | 1.09 | 0.90 |
| 1.53 | 265 | 246 | 253 | 1.04 | 1.31 | 1.20 |
| 1.70 | 258 | 238 | 246 | 1.13 | 1.44 | 1.30 |
| 1.95 | 242 | 226 | | 1.36 | 1.69 | |

For most of the samples we find very good agreement between the determined volume fractions from structure factor analysis (SAXS) and diffraction analysis by UV-Vis spectroscopy. Only in some cases we observe a mismatch with more than 10% deviation. The mismatch is highest for the higher volume fractions and is attributed to a less precise analysis of the weak (022) structure factor peak in SAXS. The volume fractions obtained from UV-Vis spectroscopy are defective to the average refractive index of the dense samples. Deviations to the SAXS data might be explained by the fact that we did not consider the concentration dependence of the refractive index. Nevertheless, the good agreement between the data for most of the samples, highlights the robustness of our structure factor analysis.

**Figure S14** shows the experimentally determined Bragg peak positions ($\lambda_{Bragg}$) from UV-Vis spectroscopy plotted as a function of $\phi^{-1/3}$. A linear fit (solid line) describes the experimental data well indicating that all samples contain crystalline domains with the same crystal structure.

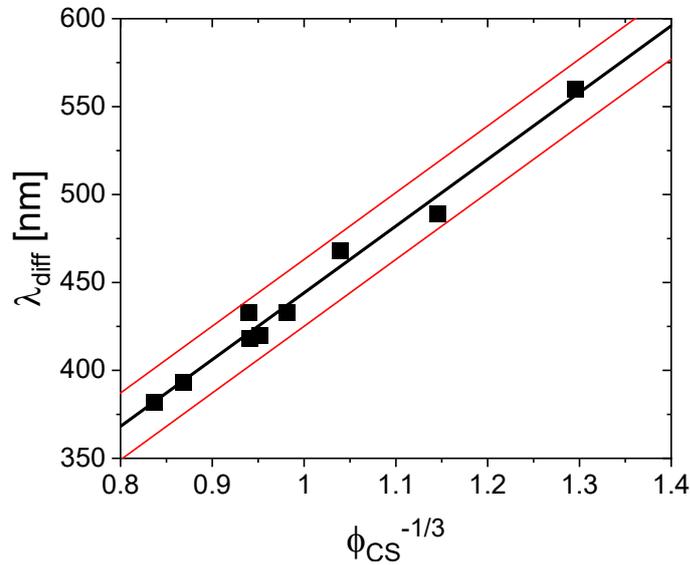

**Figure S14.** Results from Bragg peak analysis by UV-Vis spectroscopy. Plotted are the positions of the Bragg peaks ($\lambda_{Bragg}$) as function of $\phi^{-1/3}$. The black line represents a linear fit while the red lines show the respective error range.

The position of $\lambda_{Bragg}$ depends on the lattice spacing ($d_{hkl}$), the refractive index of the crystalline sample ($n_{crystal}$) and the angle between the incoming beam and the sample:

$$\lambda_{diff} = 2\, d_{hkl} \sqrt{n_{crystal}^2 - \sin\theta^2} \qquad\qquad (\textbf{S.12})$$

We prepared samples in rectangular capillaries that were studied in standard transmission geometry. For an fcc (or rhcp) crystal structure, we expect the 111 plane to be parallel to the walls of the capillary. Therefore, the 111 plane is orthogonal to the beam in transmission geometry ($\theta = 0°$). Therefore, equation **S.12** simplifies to:

$$\lambda_{diff} = 2\, d_{hkl}\, n_{crystal} \qquad\qquad (\textbf{S.13})$$

The lattice constant $a$ and the lattice spacing $d_{hkl}$ are related by the following equation:

$$a = d_{\mathrm{hkl}} \sqrt{h^2 + k^2 + l^2} \tag{S.14}$$

Combining equations **S.13** and **S.14** we get:

$$a = \frac{\lambda_{\mathrm{diff}} \sqrt{h^2 + k^2 + l^2}}{2 \, n_{\mathrm{crystal}}} \tag{S.15}$$

The volume fraction of an fcc crystal with 3 + 1 spheres per unit cell can be calculated with the lattice constant $a$ and the radius of the spheres $R$.

$$\phi = \frac{(3+1)\frac{4}{3}\pi R^3}{a^3} \tag{S.16}$$

Using $a$ according to equation **S.15**, we get:

$$\phi = \frac{(3+1)\frac{4}{3}\pi R^3}{\left(\frac{\lambda_{\mathrm{diff}} \sqrt{h^2+k^2+l^2}}{2 \, n_{\mathrm{crystal}}}\right)^3} \tag{S.17}$$

With equation **S.17** providing the relation between the volume fraction and the position of the diffraction peak, we can directly show the linear dependency between $\lambda_{\mathrm{diff}}$ and $\phi_{\mathrm{cs}}^{-1/3}$ as experimentally verified in **Figure S14**:

$$\left(\frac{\lambda_{\mathrm{diff}} \sqrt{h^2+k^2+l^2}}{2 \, n_{\mathrm{crystal}}}\right)^3 = \frac{(3+1)\frac{4}{3} R^3}{\phi} \tag{S.18}$$

$$\lambda_{\mathrm{diff}} = \frac{2 \, n_{\mathrm{crystal}} \sqrt[3]{(3+1)\frac{4}{3} R^3}}{\sqrt{h^2+k^2+l^2}} \, \phi^{-\frac{1}{3}} \tag{S.19}$$

Regarding the scaling of $\phi_{\mathrm{cs}}$ with the lattice constant $a$ shown in **Figure 4c** in the main manuscript, we can rewrite equation **S.16** in the following way:

$$a^3 = \frac{16}{3}\pi R_{\mathrm{h}}^3 \frac{1}{\phi} \tag{S.20}$$

$$a = \sqrt[3]{\frac{\pi 16}{3}} R_{\mathrm{h}} \phi^{-\frac{1}{3}} \tag{S.21}$$

Here, we use the hydrodynamic radii ($R_{\mathrm{h}}$) of the CS microgels. According to equation **S.21** the lattice constant would approach zero for high enough volume fractions. In reality, our soft and deformable CS microgels can only be packed to a certain limit. To take this into account, we add the offset $B$ to describe our experimental findings:

$$a = \sqrt[3]{\pi \frac{16}{3}} R_\mathrm{h} \phi^{-\frac{1}{3}} + B \qquad\qquad\qquad\qquad\qquad\qquad \text{(S.22)}$$

Equation **S.22** is used to the guide to the eye shown in **Figure 4c** in the main manuscript. We used a value of $B$ = 215 nm as offset. Regarding a fcc crystal structure this would result in a hard sphere radius of 76 nm which is close to the CS microgel radius of 77.9 nm that we determined from SAXS.

**Characterization of Crystalline Samples**

Due to the anisotropic scattering signal from dense samples, we can also apply analysis of the 2D detector images from SAXS. **Figure S15A** and **S15B** show the 2D SAXS patterns of two crystalline samples with $\phi_{cs}$ = 0.66 and $\phi_{cs}$ = 1.20. Both patterns show at least two orders of sharp Bragg peaks with six-fold symmetry indicative of hexagonally ordered planes aligned parallel to the capillary walls. In addition, a distinct amorphous ring is visible indicating the coexistence with fluid-like/disordered regions in the samples. Furthermore, some weak Bragg peaks are visible that are most likely caused by small, differently oriented crystal domains and will be excluded from further analysis.

**Figure S15C** and **S15D** show the overlay of the experimental scattering patterns with simulated ones obtained from Scatter.[8] Using an fcc crystal structure with small domain sizes[9] (~1 µm) we find qualitatively good agreement. We used a core shell model in our simulations with a homogeneous shell. This is comparable to our approach of fitting the form factor. The homogeneous shell is described by a constant scattering length density of the polymer shell in its radial density profile, which is not addressing the fuzzy sphere morphology known for microgels. The particle sizes used in the simulation were $R_\mathrm{core}$ = 6.5 nm and $R_\mathrm{total}$ = 91 nm for $\phi_{cs}$ = 0.66 and $R_\mathrm{total}$ = 83 nm for $\phi_{cs}$ = 1.20, respectively (see **Table S8**). The facts that both particle sizes are smaller than $R_\mathrm{h}$ =105 nm and that $R_\mathrm{total}$ decreases with $\phi_{cs}$ indicate that the microgels are in close contact and densely packed. The agreement between simulated and experimental scattering patterns is reasonable and points towards fcc as the crystal structure. This finding is in agreement with a previous SANS study of similar CS microgels.[2] However, we want to highlight that a precise analysis of the crystal lattice would require scattering data with many more orders of Bragg peaks.

While in our previous SANS study the main scattering contribution to $S(q)$ was related to scattering from the microgel shells, here the dominant contrast in SAXS comes from the small AuNP cores. The fact that we observe such sharp and pronounced Bragg peaks underlines that the AuNP cores are well centered in the microgels. In contrast pronounced variation in core location would lead to strong smearing of the structure factor.

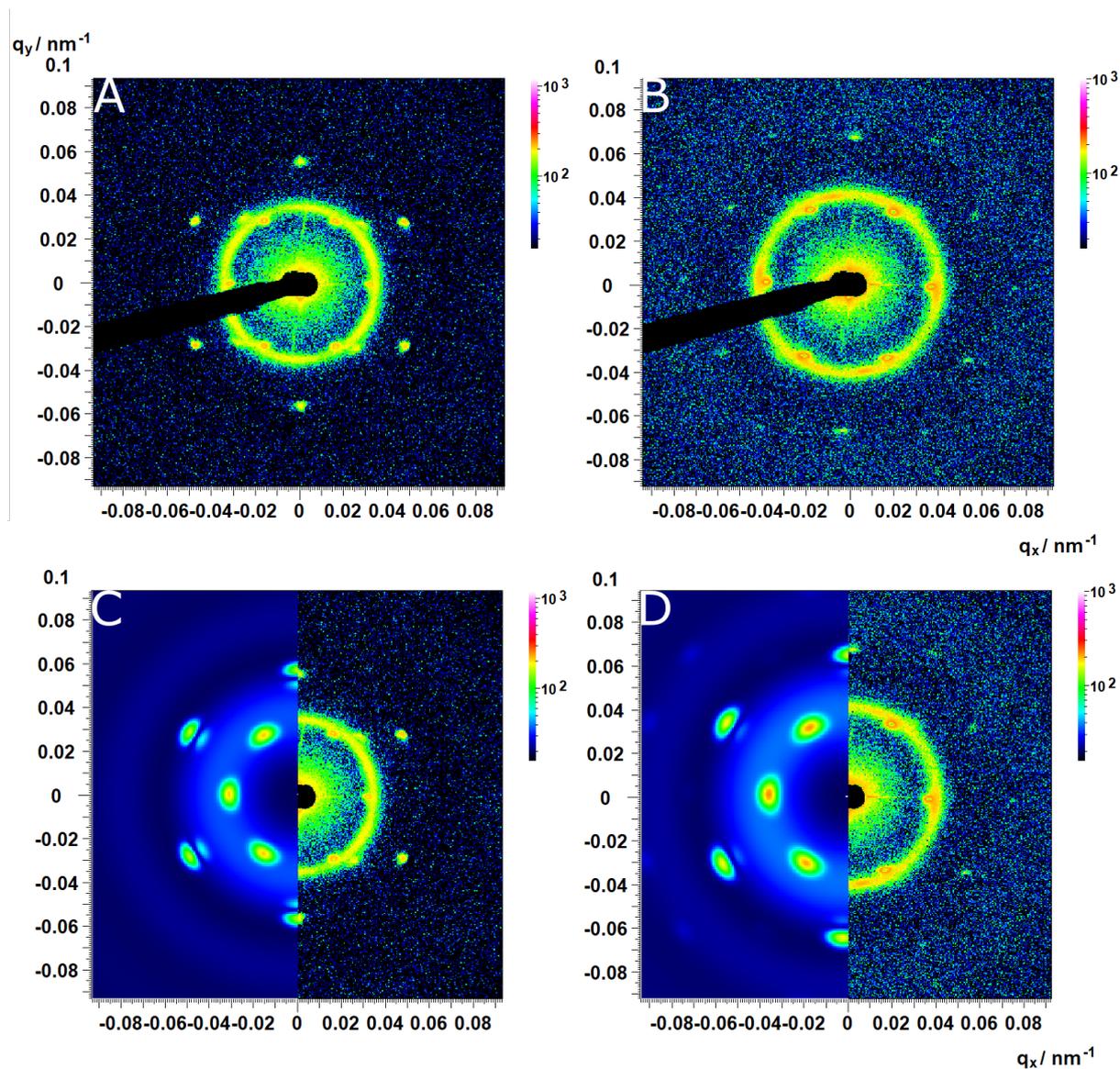

**Figure S15.** 2D SAXS patterns of the $\phi_{cs}$ = 0.66 (A) and $\phi_{cs}$ = 1.20 samples (B). (C, D) Scattering patterns simulated with Scatter software (left half) on top of the 2D SAXS patterns shown in (A) and (B) respectively.

**Table S8.** List of the parameters used for the simulation of the 2D scattering patterns of crystalline samples with the Scatter software.

| | $\phi cs = 0.66$ | $\phi cs = 1.20$ |
|---|---|---|
| **Crystal lattice** | fcc | fcc |
| Unit cell *a* [nm] | 322 | 281 |
| Radial domain [nm] | 1100 | 950 |
| Azimuthal domain [nm] | 750 | 650 |
| Maximum hkl | 2 | 3 |
| **Form Factor** | Sphere (core + homogenous shell) | Sphere (core + homogenous shell) |
| $R_{core}$ [nm] | 6.54 | 6.54 |
| $\sigma_{core}$ | 0.07 | 0.07 |
| $R_{CS}$ [nm] | 91 | 83 |
| $\rho$ (ratio between SLDs) | 0.01 | 0.01 |